\documentclass{elsart0}  

\usepackage{amsmath,amssymb}

\begin{document}

\begin{frontmatter}



\title{Field equations in teleparallel spacetime:\\
Einstein's {\it Fernparallelismus} approach towards unified field
theory.}


\author{Tilman Sauer}
\ead{tilman@einstein.caltech.edu}

\address{Einstein Papers Project\\
California Institute of Technology 20-7\\
Pasadena, CA~91125, USA}

\begin{abstract}

A historical account of Einstein's {\it Fernparallelismus}
approach towards a unified field theory of gravitation and
electromagnetism is given. In this theory, a space-time
characterized by a curvature-free connection in conjunction with a
metric tensor field, both defined in terms of a dynamical tetrad
field, is investigated. The approach was pursued by Einstein in a
number of publications that appeared in the period from summer
1928 until spring 1931. In the historical analysis special
attention is given to the question of how Einstein tried to find
field equations for the tetrads. We claim that it was the failure
to find and justify a uniquely determined set of acceptable field
equations which eventually led to Einstein's abandoning of this
approach. We comment on some historical and systematic
similarities between the {\it Fernparallelismus} episode and the
{\it Entwurf} theory, i.e.\ the precursor theory of general
relativity pursued by Einstein in the years 1912--1915.

\end{abstract}


\end{frontmatter}



\section{Introduction}

Einstein's attempt to base a unified theory of the gravitational
and electromagnetical fields on the mathematical structure of
distant parallelism, also referred to as absolute or
teleparallelism,\footnote{In this paper, these terms will be used
interchangeably.} is an episode that lasted for three years, from
summer 1928 until spring 1931. The crucial new concept, for
Einstein, that initiated the approach was the introduction of the
tetrad field, i.e.\ a field of orthonormal bases of the tangent
spaces at each point of the four-dimensional manifold. The tetrad
field was introduced to allow the distant comparison of the
direction of tangent vectors at different points of the manifold,
hence the name distant parallelism. From the point of view of a
unified theory, the specification of the four tetrad vectors at
each point involves the specification of sixteen components
instead of only ten for the symmetric metric tensor. The idea then
was to exploit the additional degrees of freedom to accommodate
the electromagnetic field. Mathematically, the tetrad field easily
allows the conceptualization of more general linear affine
connections, in particular, non-symmetric connections of vanishing
curvature but non-vanishing torsion. Since, however, Einstein
wanted to combine a curvature-free connection with a non-trivial
metric the resulting structure actually involves two different
connections and a certain ambiguity was inherent in their
interpretation.

The published record of the distant parallelism episode comprises
eight papers in the {\it
  Sitzungsberichte} of the Prussian Academy, at the time Einstein's major
forum for publication of scientific results. A review paper on the
theory appeared in the {\it Mathematische Annalen}, a leading
mathematics journal in an issue together with a historical essay
on the subject matter by Elie Cartan. The theory of distant
parallelism is also touched upon in popular articles by Einstein
for the New York and London Times. Attempts to place his new
attempt into a larger tradition of field theoretic attempts in the
history of physics are made in a contribution to a {\it
Festschrift} for Aurel Stodola and in three popular papers on the
{\it Raum-, Feld-, und \"Ather-Problem} in physics. Two of the
papers, including the very last one, were coauthored with Walter
Mayer, Einstein's mathematical collaborator. The episode is, of
course, also reflected in Einstein's contemporary correspondence,
notably with Herman M\"untz, Roland Weitzenb\"ock, Cornelius
Lanczos, Elie Cartan, and Walther Mayer. For most of the published
papers manuscript versions are extant in the Einstein Archives and
there are also a number of unidentified and undated research
calculations that are related to the distant parallelism
approach.\footnote{See especially the documents with archive
numbers 62-001ff. A thorough investigation of the unpublished
correspondence and scientific manuscripts remains to be done. In
this paper, I will largely rely on the published record of the
episode.}

As far as Einstein was involved in it, the {\it Fernparallelismus}
approach has a distinct beginning, a period of intense
investigation, and a somewhat less distinct but definite end. The
mathematical structure in question had been developed before by
others, notably by Elie Cartan and Roland Weitzenboeck, in other,
purely mathematical contexts.\footnote{I would like to emphasize
that I will refrain from any claims regarding questions of
mathematical priority with respect to the concepts of differential
geometry relevant for the episode.} Einstein's pursuit of the
approach also triggered a more general discussion that involved
quite a few other contemporary physicists and
mathematicians,\footnote{See
\cite[pp.~120ff.]{GoldsteinRitter2003}.} and it continued to be
investigated by others even when Einstein no longer took active
part in these discussions. Even today, teleparallelism is
occasionally discussed, e.g.\ as a rather special case in a more
general conceptual framework of a metric-affine gauge theory of
gravity.\footnote{See, e.g.\ \cite{Gronwald1997}, esp.~pp.~295ff;
see also \cite{Hehletal1980} which list a number of references
that have taken up the distant parallelism approach over time on
p.~341.}

In this paper, a historical account of the {\it Fernparallelismus}
approach is given as an episode in Einstein's intellectual
life.\footnote{For earlier historical discussions of the {\it
Fernparallelismus} approach, see \cite[pp.~60--67]{Treder1971},
\cite[pp.~344--347]{Pais1982}, \cite{Biezunski1989},
\cite[pp.~234--257]{Vizgin1994},
\cite[pp.~120--133]{GoldsteinRitter2003},
\cite[pp.~57--58]{vanDongen2002}, \cite[sec.~4.3.3]{Goenner2002}.
See also \cite[sec.~6.4.]{Goenner2004}, which was published after
submission of this paper.} The account will largely be organized
chronologically and give the relevant biographical data, as it
were, of the life cycle of this approach, as far as Einstein is
concerned. The infancy of the approach, sec.~\ref{sec:infancy}, is
given by Einstein's first two notes on distant parallelism which
lay out the mathematical structure and give a first derivation of
a set of field equations. I will give a brief characterization of
the approach in modern terms in sec.~\ref{sec:infancy1} and
discuss Einstein's first notes in sec.~\ref{sec:infancy2}. In its
early childhood, Einstein entered into interaction with
mathematicians and learnt about earlier pertinent developments in
mathematics , sec.~\ref{sec:childhood}. I will briefly comment on
his correspondence with Herman M\"untz, sec.~\ref{sec:childhood1},
and Roland Weitzenb\"ock, sec.~\ref{sec:childhood2}, as well as on
his collaboration with Jakob Grommer and Cornelius Lanczos,
sec.~\ref{sec:childhood3}. The period of adolescence is primarily
concerned with the problem of finding field equations,
sec.~\ref{sec:adolescence}. I first discuss the publication
context of Einstein's next papers in sec.~\ref{sec:adolescence1}
and then focus on his attempts to find and justify a set of field
equations in sec.~\ref{sec:adolescence2}. Einstein here wavered
between a variational approach, sec.~\ref{sec:adolescence2a}, and
an approach where field equations were determined utilizing
algebraic identities for an overdetermination of the equations,
sec.~\ref{sec:adolescence2b}. The mature stage is reached when
Einstein settled on a set of field equations and wrote an overview
of the theory published in the {\it Mathematische Annalen},
sec.~{\ref{sec:maturity}. I will first give an account of the
publication history of this paper which is intimately linked with
Einstein's correspondence with Elie Cartan,
sec.~\ref{sec:maturity1}, and then discuss the derivation of the
field equations, sec.~\ref{sec:maturity2}. In its old age,
sec.~\ref{sec:oldage}, Einstein improved on the compatibility
proof, sec.~\ref{sec:oldage1}, promoted the theory in public and
defended it against criticism, sec.~\ref{sec:oldage2}, and
explored its consequences, sec.~\ref{sec:oldage3}. The last sign
of life is a paper that systematically investigated compatible
field equations in a teleparallel space-time,
sec.~\ref{sec:oldage4}.

The life cycle of the distant parallelism approach bears a number
of striking similarities to the life cycle of the {\it Entwurf}
theory, i.e.\ the precursor theory of general relativity advanced
and pursued by Einstein between 1912 and 1915 and presented first
in an {\it Outline} (``Entwurf'') {\it of a Generalized Theory of
Relativity and a Theory of Gravitation} in 1913
\cite{EinsteinGrossmann1913}. Some of these similarities between
the history of the {\it Fernparallelismus} approach and the {\it
Entwurf} theory will be pointed out along the way. They are, I
believe, no coincidence. I will offer some reflections on the
systematic reason for this similarity in the concluding remarks in
sec.~\ref{sec:conclusion}.

\section{Einstein's distant parallelism as a mathematical structure}
\label{sec:infancy}

Before entering into the discussion of the historical material,
the mathematical framework of Einstein's distant parallelism shall
here roughly be characterized in modern terms. I will then discuss
Einstein's first two notes on the subject.

\subsection{Modern characterization}
\label{sec:infancy1}

The basic ingredients are a bare differentiable manifold, a
curvature-free connection that allows to define a frame field on
the tangent bundle, and the demand of global $SO(n-1,1)$-symmetry
that allows to define a metric tensor field in a meaningful way.
Naturally, a coordinate-free characterization raises issues of
global existence and similar concerns which, however, I will not
discuss here.

\paragraph*{Step 0.}

The starting point is an $n$-dimensional real, differentiable,
$C^{\infty}$-manifold $M$ just as in any modern account of the
mathematical structure of general relativity, i.e.\ if needed one
might specify it as paracompact, Haussdorff, etc.

\paragraph*{Step 1.}

Let  $\vartheta_a$ be frame field on $M$, i.e.\ a set of $n$
linearly independent, differentiable vector fields or, in other
words, a cross section of the frame bundle. Such a frame field may
not exist globally. If that is the case, we restrict ourselves to
a parallelizable subset of $M$. At this point, $\vartheta_a$ is
not specified. It will be obtained later as a solution to some set
of field equations.

We now introduce a a connection, i.e. a
$\operatorname{gl}(\Bbb{R},n)$-valued one-form $\omega^a_b$ on the
tangent bundle $TM$ that is compatible with $\vartheta_a$, in the
sense that the associated parallel transport is realized by the
field $\vartheta_a$, i.e.\ the covariant derivative of the frame
vectors vanishes.

This condition determines the connection uniquely. By patching
together information from different coordinate charts, it can also
be defined globally, even if a global frame field does not exist.
Without historical prejudice, we shall call the connection the
Weitzenb\"ock connection. The curvature form $O^a_b = d\omega^a_b
+ \omega^a_m \wedge \omega ^m_b$ for this connection vanishes
$O^a_b \equiv 0$. Its torsion two-form $\Omega_a=d\vartheta_a -
\vartheta_m \wedge \omega^m_a$, however, does not vanish in
general. Conversely, a Weitzenb\"ock connection $\omega^a_b$ does
not uniquely determine a frame field $\vartheta_a$. The frame
field is only determined up to a {\it global}
$GL(\Bbb{R},n)$-transformation.

The global $GL(\Bbb{R},n)$-symmetry can also be seen like this.
Given a Weitzenb\"ock connection over $M$ and a local frame at a
single point $p\in M$, we can parallel transport the frame over
the tangent bundle and construct a frame field $\vartheta_a$.
Obviously, we could start with any linearly independent set of
vectors in $T_{p}M$ and would obtain different frame fields for
each such frame in $T_pM$ which are equivalent up to global
Lorentz rotations.

\paragraph*{Step 2.}

Given the frame field, we can now define a metric by conceiving of
the frame field as an orthonormal vector field. To be specific, we
will now assume the manifold to be of dimension $n=4$. We would
then obtain a metric tensor field by
\begin{equation}
g = o^{ab}\vartheta_a\vartheta_b.
\end{equation}
where
\begin{equation}
o^{ab} = \operatorname{diag}(-1,-1,-1,+1).
\end{equation}
The definition of the metric reduces the global
$GL(\Bbb{R},4)$-symmetry to a global $SO(3,1)$-symmetry. This
symmetry requirement defines Einstein's version of distant
parallelism. Any frame field $\vartheta_a$ uniquely defines a
metric tensor field. The converse is not true, since a metric
tensor field is determined by $n(n+1)/2$ components, whereas a set
of $n$ linear independent vectors is defined by $n^2$ components.
In $n=4$ dimensions, the metric tensor determines $10$ of the $16$
components leaving $6$ components undetermined. This is just the
amount of freedom needed to accommodate the electromagnetic field
in the theory.

The existence of a metric tensor field on $M$ allows the
definition of a second, uniquely defined, metric compatible
connection, i.e.\ the usual Levi-Civita connection. Its torsion
two-form vanishes while its curvature two-form in general does not
vanish. The curvature associated with the Levi-Civita connection
vanishes if and only if the Weitzenb\"ock torsion
vanishes.\footnote{This relativizes Pais's observation, that the
distant parallelism approach was unusual for Einstein because
``the most essential feature of the `old' theory is lost from the
very outset: the existence of a nonvanishing curvature tensor.''
\cite[p.~344]{Pais1982}. This remark ignores the fact that more
than one connection plays a role in this theory.}

These then are the basic ingredients of Einstein's distant
parallelism: A curvature free Weitzenb\"ock connection and the
demand of global $SO(n-1,1)$-symmetry. The Weitzenb\"ock
connection defines a frame field, and the global rotation symmetry
assures that the frame field can determine a metric tensor field
in a meaningful way. But since a given metric field would not
uniquely determine a frame field, Einstein had introduced a
surplus structure which he hoped to be able to exploit for setting
up a unified field theory of the gravitational and electromagnetic
fields.

In 1912, it was the metric tensor that opened up new possibilities
for exploring a generalized theory of relativity and a field
theory of gravitation. Similarly new vistas had been opened by
taking the concept of a (symmetric) connection as the new basic
mathematical ingredient. Now it was the introduction of a tetrad
field that provided new possibilities as well as new constraints.

Since two distinct connections are involved in the structure, a
certain ambiguity is involved as to which connection is the
physically meaningful one. But this ambiguity may not become
explicit in a unified theory, since no external matter fields are
assumed for which one would have to decide which connection should
determine covariant differentiation. Also it has not been
determined how the electromagnetic field is to be defined in terms
of the frame field. Finally, in order to set up a physically
meaningful structure the frame field needs to be determined by
some set of field equations.

\subsection{Einstein's first two notes}
\label{sec:infancy2}

The episode of {\it Fernparallelismus}, as far as Einstein is
concerned, begins with two rather short notes, 5 and 4 pages each,
published within a week's interval in the {\it Sitzungsberichte}
of the Prussian Academy. The first note is entitled {\it
Riemannian geometry, maintaining the concept of distant
parallelism} \cite{Einstein1928a} and was presented to the Academy
on June 7, 1928.

Einstein at the time suffered from a serious illness of his
heart.\footnote{For the following biographical information, see
\cite[pp.~600--607]{Foelsing1997}.} He had experienced a
circulatory collapse in Switzerland in March. An enlargement of
the heart was diagnosed and, back in Berlin, he was ordered strict
bed rest as well as a salt-free diet and diuretics. At the end of
May, he wrote to his friend Zangger: ``In the tranquility of my
sickness, I have laid a wonderful egg in the area of general
relativity. Whether the bird that will hatch from it will be vital
and long-lived only the Gods know. So far I am blessing my
sickness that has endowed me with it.''\footnote{``Ich habe in der
Ruhe der Krankheit ein wundervolles Ei gelegt auf dem Gebiete der
allgemeinen Relativit\"at. Ob der daraus schl\"upfende Vogel vital
und langlebig sein wird, liegt noch im Schosse der G\"otter.
Einstweilen segne ich die Krankheit, die mich so begnadet hat.''
Einstein to Zangger, end of May 1928, Einstein Archives, The
Hebrew University of Jerusalem (EA), call no. 40-069.} Since he
was feeling too weak to attend the Academy meetings, his note was
presented to the Academy by Max Planck.

The paper explains the notion of a tetrad field
(``$n$-Bein''-field) and of distant parallelism
(``Fernparallelismus'') for a manifold of $n$-dimensions. The
tetrad field is introduced in terms of components $h_s{}^{\nu}$ of
its vectors with respect to the naturally induced coordinate
basis. Hence $h_s{}^{\nu}$ denotes the $\nu$-component of the
vector $s$ with respect to the local coordinate chart. Einstein
uses greek letters to denote the coordinate indices
(``Koordinaten-Indizes'') and latin letters to denote the tetrad
indices (``Bein-Indizes''). In modern literature, these indices
are also referred to as holonomic resp.\ anholonomic. We have the
relations
\begin{align}
h_{a\mu}h^{a\nu} &= \delta^{\nu}_{\mu}, \\
h_{a\mu}h_b{}^{\mu} &= \delta_{ab}, \label{eq:delta_ab}
\end{align}
where a summation over repeated indices is always
implied.\footnote{Einstein's original notation did not distinguish
between tetrad vectors and the canonical dual covectors, i.e.\ he
did not use superscripted tetrad indices. In general, I will not
adhere strictly to Einstein's original notation. In particular, I
will denote partial coordinate derivatives by comma-delimited
subscripts.}

Einstein emphasized that the tetrads define both the metric and
the distant parallelism simultaneously:
\begin{quote}
By means of the introduction of the $n$-Bein field both the
existence of a Riemann-metric and the existence of the distant
parallelism is expressed.\footnote{``Durch die Setzung des
$n$-Bein-Feldes wird gleichzeitig die Existenz der Riemann-Metrik
und des Fernparallelismus zum Ausdruck gebracht.''
\cite[p.~218]{Einstein1928a}.}
\end{quote}
The components of the metric tensor $g_{\mu\nu}$ are given as
\begin{equation}
g_{\mu\nu} = h^a{}_{\mu}h_{a\nu} \label{eq:gintermsofh}.
\end{equation}
By virtue of (\ref{eq:gintermsofh}) coordinate indices are raised
and lowered using the metric $g_{\mu\nu}$, whereas by
(\ref{eq:delta_ab}) tetrad indices are raised and lowered using
$\delta_{ab}$.

Parallel transport is defined through the tetrads, in the sense
that a vector with components $A_a$ at one point shall be parallel
to a vector ${A'}_a$ at another point if the components with
respect to the respective tetrads are the same. The law of
parallel transport is hence given by the condition
\begin{equation}
0 = dA_a = d(h_{a\mu}A^{\mu}) = h_{a\mu,\sigma}A^{\mu}dx^{\sigma}
+ h_{a\mu}dA^{\mu}.
\end{equation}
Multiplication with $h^{a\nu}$ turns this into
\begin{equation}
dA^{\nu} = -\Delta^{\nu}_{\mu\sigma} A^{\mu}dx^{\sigma},
\end{equation}
where the connection
\begin{equation}
\Delta^{\nu}_{\mu\sigma} = h^{a\nu}h_{a\mu,\sigma}
\label{eq:Weitzconn}
\end{equation}
is introduced.\footnote{As pointed out already by
\cite[p.~687]{Reichenbach1929} Einstein's original paper contained
several typographical errors in these equations, see also
\cite[pp.~121]{GoldsteinRitter2003}.} As Einstein noted it is
``rotation invariant'' and asymmetric in its lower indices.
Parallel transport along a closed line reproduces the same vector,
i.e.\ the Riemann curvature,
\begin{equation}
R^{\iota}_{\kappa,\lambda\mu} = -
\Delta^{\iota}_{\kappa\lambda,\mu} +
\Delta^{\iota}_{\kappa\mu,\lambda} +
\Delta^{\iota}_{\alpha\lambda}\Delta^{\alpha}_{\kappa\mu}
-\Delta^{\iota}_{\alpha\mu}\Delta^{\alpha}_{\kappa\lambda} \equiv
0, \label{eq:Riemanncurvature}
\end{equation}
vanishes identically.

Einstein observed that the metric (\ref{eq:gintermsofh}) gives
rise to another, non-integrable law of parallel transport, that is
determined by the symmetric Levi-Civita connection,
\begin{equation}
\Gamma^{\nu}_{\mu\sigma} = \frac{1}{2}g^{\nu\alpha} \left(
g_{\mu\alpha,\sigma} + g_{\sigma\alpha,\mu} - g_{\mu\sigma,\alpha}
\right).
\end{equation}
He also introduced the contorsion tensor
$\Gamma^{\nu}_{\alpha\beta}-\Delta^{\nu}_{\alpha\beta}$, and the
torsion tensor,
\begin{align}
\Lambda^{\nu}_{\alpha\beta} &= \frac{1}{2}\left(
\Delta^{\nu}_{\alpha\beta}-\Delta^{\nu}_{\beta\alpha} \right) \\
&=
\frac{1}{2}h^{a\nu}\left(h_{a\alpha,\beta}-h_{a\beta,\alpha}\right),
\label{eq:Lambda}
\end{align}
although he does not use those names for these quantities.

The possibility of obtaining field equations from a variational
principle,
\begin{equation}
\delta\int\left\{\mathcal{H}d\tau\right\} = 0,
\label{eq:varpinciple}
\end{equation}
is briefly indicated. The variation would have to be done with
respect to the the sixteen quantities $h_{a\mu}$ and the
Lagrangian $\mathcal{H}$ would have to be a linear function of the
two invariants
$g^{\mu\nu}\Lambda^{\alpha}_{\mu\beta}\Lambda^{\beta}_{\nu\alpha}$
and $g_{\mu\nu} g^{\alpha\sigma} g^{\beta\tau}
\Lambda^{\mu}_{\alpha\beta}\Lambda^{\nu}_{\sigma\tau}$, multiplied
with the determinant $h=|h_{a\mu}|$ since $hd\tau$ is an invariant
volume element.

The second note is entitled ``New possibility for a unified field
theory of gravitation and electricity'' \cite{Einstein1928b} and
was presented to the Academy only a week after the first paper, on
14 June 1928. In the introduction, Einstein wrote that it had
occurred to him in the meantime that the structure of distant
parallelism allows the identification of the gravitational and
electromagnetic field equations in a most natural manner. He
specialized to the case of four dimensions and identified the
electromagnetic potential with the quantity
\begin{equation}
\phi_{\mu}\equiv\Lambda^{\alpha}_{\mu\alpha} =
\frac{1}{2}h^{a\nu}\left(h_{a\mu,\nu}-h_{a\nu, \mu}\right).
\label{eq:phiintermsofh}
\end{equation}
More precisely, he stated that $\phi_{\mu} = 0$ would be the
mathematical expression for the absence of any electromagnetic
field. But he added in a footnote that the same could be expressed
by the condition $\phi_{(\mu,\nu)}=0$ and observed that this fact
would result in a ``certain indeterminateness of the
interpretation'' (``gewisse Unbestimmtheit der Deutung.'').

The field equations are now given by specifying the Lagrangian
$\mathcal{H}$ as
\begin{align}
{\mathcal H} &= h
g^{\mu\nu}\Lambda^{\alpha}_{\mu\beta}\Lambda^{\beta}_{\nu\alpha}
\label{eq:Rx175L} \\
&= \frac{1}{4}hh^a{}_{\mu}h_{a\nu}
h^{b\alpha}\left(h_{b\mu,\beta}-h_{b\beta,\mu}\right)
h^{c\beta}\left(h_{c\nu,\alpha}-h_{c\alpha,\nu}\right).
\end{align}
In linear approximation, $h_{a\mu}=\delta_{a\mu} +
\tilde{h}_{a\mu}$, $|\tilde{h}_{a\mu}|, |\partial \tilde{h}_{a\mu}|\ll 1$,
Einstein obtained the field equations explicitly as
\begin{equation}
\tilde{h}_{\beta\alpha,\mu\mu} - \tilde{h}_{\mu\alpha,\mu\beta} +
\tilde{h}_{\alpha\mu,\mu\beta} - \tilde{h}_{\beta\mu,\mu\alpha} = 0.
\label{eq:linearizedfe}
\end{equation}
Introducing the metric field in first approximation as
\begin{equation}
\tilde{g}_{\alpha\beta} = \delta_{\mu\nu} +
\tilde{h}_{\alpha\beta} + \tilde{h}_{\beta\alpha},
\end{equation}
and the electromagnetic four-potential $\tilde{\phi}_{a}$ as
\begin{equation}
\tilde{\phi}_{a} = \frac{1}{2}\left(\tilde{h}^{\mu}_{\alpha,\mu} -
\tilde{h}^{\mu}_{\mu,\alpha}\right),
\end{equation}
the linearized field equations (\ref{eq:linearizedfe}) turn into
\begin{equation}
\frac{1}{2}\left(-\tilde{g}_{\beta\alpha,\mu\mu} -
\tilde{g}_{\mu\alpha,\mu\beta} + \tilde{g}_{\alpha\mu,\mu\beta} -
\tilde{g}_{\beta\mu,\mu\alpha}\right) =
\tilde{\phi}_{\alpha,\beta}-\tilde{\phi}_{\beta,\alpha}.
\label{eq:linearizedfe2}
\end{equation}
Since the absence of any electromagnetic field was expressed by
$\phi_{\mu}\equiv 0$, (\ref{eq:linearizedfe2}) then turns into the
linear approximation of the Ricci tensor $R_{\alpha\beta}$, given
in terms of the metric, just as in standard general relativity.

The vacuum Maxwell equations are recovered in this approximation
by taking the divergence of $\tilde{\phi}_{\alpha}$ which vanishes
on account of (\ref{eq:linearizedfe2}) contracted over $\alpha$
and $\beta$, which gives
\begin{equation}
\tilde{\phi}_{\alpha,\alpha} = 0, \label{eq:linMaxw1}
\end{equation}
and by
\begin{equation}
\tilde{\phi}_{\alpha,\beta\beta} = 0, \label{eq:linMaxw2}
\end{equation}
which follows from the fact that the left hand side
$L_{\alpha\beta}$ of (\ref{eq:linearizedfe2}) satisfies the
identity
\begin{equation}
\left(L_{\alpha\beta} -
\frac{1}{2}\delta_{\alpha\beta}L_{\sigma\sigma}\right)_{,\beta}
=0.
\end{equation}
Eqs.~(\ref{eq:linMaxw1}) and (\ref{eq:linMaxw2}) together imply
the vanishing of the divergence of the electromagnetic field
$\phi_{\mu,\nu}-\phi_{\nu,\mu}$ which is just the inhomogeneous
set of Maxwell equations in the absence of an external current.
The homogeneous Maxwell equations are, of course, trivially
fulfilled if an electromagnetic potential exists.

In a note added at proof stage, he observed that quite similar
results could be obtained for the Lagrangian
\begin{equation}
{\mathcal H} = h g_{\mu\nu} g^{\alpha\sigma} g^{\beta\tau}
             \Lambda^{\mu}_{\alpha\beta}\Lambda^{\nu}_{\sigma\tau}
             \label{eq:Rx175Ladd}
\end{equation}
and concluded that there is an ambiguity in the choice of
${\mathcal H}$.

\section{Interaction with others}
\label{sec:childhood}

Einstein's first two notes on teleparallelism appear to be
conceived and composed without any interaction with other
mathematicians or physicists. This is confirmed by Einstein
explicitly.
\begin{quote}
After twelve years of searching with many disappointments I now
discovered a metric continuum structure that lies between the
Riemannian and the Euclidean structures and the elaboration of
which leads to a truly unified field theory.\footnote{``Nach
zw\"olf Jahren entt\"auschungsreichen Suchens entdeckte ich nun
eine metrische Kontinuumstruktur, welche zwischen der Riemannschen
und der Euklidischen liegt, und deren Ausarbeitung zu einer
wirklich einheitlichen Feldtheorie f\"uhrt.''
\cite[p.~130]{Einstein1929a}.}
\end{quote}
Nor does Einstein acknowledge any relevant literature in those
first two notes. The only reference to existing work in the field
that he did give concerned a --- problematic --- comparison of the
{\it
  Fernparallelismus} approach with the standard Riemannian geometry
and with Weyl's {\it Nahegeometrie}. Pointing out that in Weyl's
{\it Infinitesimalgeometrie} parallel transport would preserve
neither lengths nor directions of vectors he puts his own theory
in parallel to Riemannian geometry. The latter allowed the
comparison of lengths over finite distances, but not directions,
while the former allowed parallel transport of directions but not
of lengths. The comparison is problematic because from a modern
point of view, it would seem more natural to parallelize the {\it
Fernparallelismus} to Weyl's theory as two different ways of
generalizing the underlying connection.\footnote{See
\cite{Reichenbach1929} and \cite[p.~121]{GoldsteinRitter2003}, for
  further discussion of this point.} In any case, the reference is too
vague to be counted as a real citation.

Soon after the publication of Einstein's first two notes this
situation changed. Einstein entered into intense interaction with
several other mathematicians and scientists. He began a
collaboration with the mathematician Herman M\"untz on special
solutions of the theory. He was alerted to earlier pertinent work
in the mathematics literature by Roland Weitzenb\"ock. And later
in the year, Cornelius Lanczos joined Einstein in Berlin on a year
of absence from Frankfurt. He also acknowledged contributions by
Jakob Grommer who had been working with him in Berlin all the
time. Interactions with Elie Cartan and Walther Mayer were also
important but will be discussed later on since they began much
later.

\subsection{The correspondence with Herman M\"untz}
\label{sec:childhood1}

Chaim Herman M\"untz (1884--1956) had studied mathematics in Berlin
and had obtained his Ph.D.\ in 1910 with a thesis on the partial
differential equations of the minimal surface.\footnote{For
  biographical information on M\"untz, see \cite[pp.~491f]{Pais1982}
   and \cite{PinkusOrtiz}.} The existing correspondence between
   Einstein and M\"untz in the
Einstein Archives suggest that M\"untz and Einstein had contact
already before summer 1928. M\"untz was living in Berlin at the
time, and according to Pinkus and Ortiz \cite{PinkusOrtiz} he may
have been working as Einstein's scientific collaborator as early
as summer 1927.

Their extensive correspondence about teleparallelism then appears
to have been triggered by a letter from M\"untz in which he
pointed out that the field equations in
first approximations are fully integrable.%
\footnote{``Es handelt sich darum, dass man die ersten
N\"aherungsgleichungen [...] vollst\"andig integrieren kann.''
Einstein to M\"untz, 26 July 1928, EA~18-328.}
In the sequel, M\"untz was concerned with the task of computing
the special case of spatial spherical symmetry. The correspondence
shows that Einstein kept M\"untz informed about his considerations
regarding the proper field equations, asking him about explicit
calculations for each new version of them. These calculations are
acknowledged in \cite[p.~132]{Einstein1929a} and in
\cite[p.~7]{Einstein1929b}. M\"untz was also credited with
pointing out the problem of compatibility of the field equations
derived in \cite{Einstein1929b}, see \cite[p.~156]{Einstein1929c}.
In fact, in a letter, dated 18 March 1929 (EA~18-355), M\"untz
suggested rewriting an earlier version of the introduction of
\cite{Einstein1929c}. Their collaboration ended some time in 1929
when M\"untz accepted a call as professor of mathematics at the
university of Leningrad.

\subsection{The correspondence with Roland Weitzenb\"ock}
\label{sec:childhood2}

A few days after Einstein had learnt from M\"untz about the
possibility to find explicit solutions for his equations in first
approximation, he received further correspondence regarding his new
theory. On August 1, 1929, Einstein received a letter saying:
\begin{quote}
The connection components that you denote [...]
by $\Delta^{\nu}_{\mu\sigma}$ were published first (1921) in my
encyclopedia article III E 1 in note 59 with No 18; more
explicitly in my invariant theory (1923) (Groningen: Noordhoff),
p.~317ff.\footnote{``Die von Ihnen
  [...] 
  $\Delta^{\nu}_{\mu\sigma}$ genannten Zusammenhangskomponenten finden
  sich zuerst (1921) in meinem Enzyklop\"adie-Artikel III E 1 in
  Anmerkung 59 bei No 18; ausf\"uhrlicher in meiner Invariantentheorie
  (1923) (Groningen: Noordhoof), p.~317ff.'' Weitzenb\"ock to
  Einstein, 1 August, 1929, EA~23-367.}
\end{quote}
The author was Roland Weitzenb\"ock (1885--1955), who had been
appointed professor of mathematics at the University of Amsterdam
in 1921 at the initiative of Brouwer
\cite[sec.~9.4]{vanDalen1999}. The references are to
\cite{Weitzenboeck1921} and \cite{Weitzenboeck1923}.

Weitzenb\"ock also listed some later papers by G.~Vitali,
G.F.C.~Gries, M.~Euwe, E.~Bortolotti and L.P.~Eisenhart\footnote{All
  references given in the letter are included in the more complete
  list given in \cite[p.~466]{Weitzenboeck1928}.} that would deal
with the issue of parallel transport and differential invariants in
manifolds endowed with an $n$-{\it Bein}-field.

More specifically, Weitzenb\"ock stated a formal result relevant
for Einstein's attempts to derive the field equations on the basis
of a variational formulation. He claimed that any Lagrangian, i.e.
any function that is invariant under both general coordinate
transformations {\it and} rotations of the tetrads, can be built
up from $h=|h_{a\nu}|$, $g_{\mu\nu}$, $g^{\mu\nu}$,
$\Lambda^{\nu}_{\alpha\beta}$ and its covariant derivatives with
respect to the connection $\Delta^{\nu}_{\alpha\beta}$.  Moreover,
he stated the proposition that $h$ is the only such function of
order zero,\footnote{The order of the function is defined to be
the highest
  order of differentiation in its arguments, see
  \cite[p.~470]{Weitzenboeck1928}.} no function of first order exist that
is linear in $\Lambda^{\nu}_{\alpha\beta}$, and any function of
first order that is quadratic in $\Lambda^{\nu}_{\alpha\beta}$ is
built up of the three invariants (see
eqs.~(\ref{eq:Winvariant1})-(\ref{eq:Winvariant3}) below).
Incidentally, these quantities are sometimes referred to as
Weitzenb\"ock invariants in modern literature. He announced that
he was going to write a short communication about these results
and asked whether Einstein would be willing to present such a note
to the Prussian Academy for publication in its proceedings.

Einstein was quick to respond on 3 August, two days later, that he had
written the first two notes while lying in bed with a ``severe heart
problem''
and that he had asked Planck to inquire from the mathematicians in
the Academy whether such notions are in fact known to the
mathematicians. However, Planck had told him that a publication
would be justified already from the physics point of view and
hence he, Einstein, had given in. Of course, he would be all in
favour of publishing a note by Weitzenb\"ock.

Einstein added that he had in the meantime lost some confidence in
the theory. While the quantities
$\phi_{\mu}=\Lambda^{\alpha}_{\mu\alpha}$ would satisfy Maxwell's
equations, one would not, conversely, have a corresponding tetrad
field for any solution of the Maxwell equations. In particular, a
spherically symmetric electric field seemed not to exist in the
new theory.

Weitzenb\"ock sent his note without further delay on August 8. In
his letter, he also asked a couple of questions about Einstein's
second note. One point concerned the Einstein's approximation
procedure and was clarified to be due to the fact that in setting
$h_{a\mu} = \delta_{a\mu} + \tilde{h}_{a\mu}$ Einstein had also,
but only tacitly assumed that the derivatives $\partial
\tilde{h}_{a\mu}$ would be of first order as well. The second
point concerned the question as to how to recover the vacuum field
equation of the old theory of general relativity from the
Weitzenb\"ock invariants.

In his response, Einstein explained his approximation
procedure.\footnote{Einstein somewhat missed, however, Weitzenb\"ock's
  point: ``Ich kann nicht begreifen, was Sie an meiner
  diesbez\"uglichen einfachen Rechnung auszusetzen haben.'' It was
  Weitzenb\"ock himself who gave the answer to his own question in his
  response letter.} He did not respond to Weitzenb\"ock's second point
of recovering the old gravitational equations\footnote{That point was
  addressed later in a letter by Lanczos, see the discussion below.}
but he reiterated his new doubts with respect to the viability of
the theory since it did not readily allow for the existence of
electrically charged particle-like solutions. But he added:
\begin{quote}
  But one has to be careful with a definite judgement since the
  limits of validity of the Maxwellian equations is an unsolved
  problem.\footnote{``Man muss aber mit einem endg\"ultigen Urteil
  vorsichtig sein, da die Grenze der G\"ultigkeit der Maxwell'schen
  Gleichungen ein ungekl\"artes Problem ist.'' Einstein to
  Weitzenb\"ock, 16 August 1928, Centre for Mathematics and Computer
  Science (Amsterdam), library.}
\end{quote}
He continued with an interesting heuristic comment indicating that
he would be prepared to call into question other aspects of his
heuristics if this should be necessary.
\begin{quote}
  In any case, the combination of an integrable parallel transport
  with a metric seems to me very natural since already the
  assumption of a metric in a single point of the continuum
  overdetermines the metric if the law of parallel transport is
  given. But the metric need not be defined by a quadratic function.
  However, this is made probable by the principle of the constancy
  of the velocity of light.\footnote{``Jedenfalls erscheint
  mir die Kombination einer integrablen
  Parallelverschiebung mit einer Metrik sehr nat\"urlich, da schon die
  Annahme der Metrik in {\it einem} Punkte des Kontinuums die Metrik
  \"uberbestimmt, wenn das Verschiebungsgesetz gegeben ist. Allerdings
  brauchte die Metrik nicht durch eine quadratische Funktion definiert
  zu sein, aber daf\"ur spricht das Prinzip von der Konstanz der
  Lichtgeschwindigkeit.'' ibid.}
\end{quote}
Einstein promised to present Weitzenb\"ock's note to the Academy
on the very next occasion. Due to the summer break, the next
meeting, however, took place only in October and Weitzenb\"ock's
note was indeed presented on October 18, and its published version
was issued on 28 November 1928.

Einstein mentioned Weitzenb\"ock in three of his next papers and
temporarily adopted his notation for the $n$-{\it Bein}s. But
their correspondence seems to have ended at this point.

\subsection{The cooperation with Jakob Grommer and Cornelius Lanczos}
\label{sec:childhood3}

The epistolary exchange with the mathematicians M\"untz and
Weitzenb\"ock had been triggered by the publication of Einstein's
first two notes. Two other scientists were important for Einstein
at this time, his long-standing assistent Jakob Grommer and the
theoretical physicist Cornelius Lanczos.

Jakob Grommer (1879--1933) had been working with Einstein for
several years.\footnote{For biographical information on Grommer,
see \cite[pp.~487f]{Pais1982}.} In fact, in 1925 Einstein wrote
that Grommer ``had faithfully assisted me in recent years with all
caclulations in the area of general relativity
theory.''\footnote{Quoted ibid.} Their collaboration resulted in a
number of joint publications. As to Grommer's role in the {\it
  Fernparallelismus} project, there is only few correspondence since
most of their interaction was in person. Grommer had voiced doubts
about the equivalence of the electromagnetic equations obtained in
linear approximation with Maxwell's equations in Einstein's first
version of field equations.\footnote{See Einstein to M\"untz, end
of July 1928, EA~18-311.} Einstein acknowledged Grommer's
assistance in \cite{Einstein1929b} but did not specify his
contribution.

Some time in 1929, Grommer seems to have gone to Minsk to accept a
teaching position at the university. Possibly in an attempt to
find a successor for Grommer, Einstein was eager to arrange for
Cornelius Lanczos (1893--1974) to come to Berlin for a year.
Lanczos was {\it
  Privatdozent} at the university of Frankfurt and took a year of
leave of absence in order to be able to work with Einstein in
Berlin.\footnote{For more biographical information on Lanczos and
an account of his interactions with Einstein, see
\cite{Stachel1994}.} Lanczos started to work with Einstein in
Berlin on November 1, 1928. The stay was supported by a grant from
the {\it Notgemeinschaft Deutscher
  Wissenschaft}. Einstein thanked Lanczos in the introduction of
\cite{Einstein1929c} for pointing out a problem with the
compatibility of the field equations in that note. Lanczos also
found out that the Lagrangian advanced in \cite{Einstein1929c} is
equivalent to the usual Riemann scalar (see the discussion below
in sec.~\ref{sec:maturity1}). Lanczos himself also published a
little semi-popular note on the {\it Fernparallelismus} theory
\cite{Lanczos1929} in July 1929 and a more extended but also
non-technical account in 1931 \cite{Lanczos1931}.

\section{Searching for field equations}
\label{sec:adolescence}

\subsection{Einstein's next papers}
\label{sec:adolescence1}

Further progress and the interaction with the aforementioned
mathematicians is reflected in a semi-popular overview of the
present state of field theory, two further notes on the subject in
the {\it Sitzungsberichte} and two newspaper articles.

Soon after Weitzenb\"ock's note appeared in late November,
Einstein had a chance to react to it in print. In early November
1928, he had been asked to contribute to a {\it Festschrift} on
the occasion of the seventieth birthday of Aurel Stodola,
professor of mechanical engineering at Zurich's polytechnic. That
birthday would take place on May 10, 1929, but the {\it
Festschrift} was to be completed ahead of time. Einstein agreed to
contribute a semi-popular review article ``On the Present State of
Field Theory'' \cite{Einstein1929a}. The manuscript for this paper
was submitted on 10 December 1928.\footnote{See marginal notes on
EA~22-261 and EA~22-262.}

At the end of this more general survey of the history of field
theory, Einstein briefly sketched his new approach, commenting
also on the derivation of field equations. He mentioned
calculations of the equations of motion for chargeless particles,
undertaken together with M\"untz. With reference to Weitzenb\"ock,
Einstein introduced a change of notation: algebra indices are now
written to the left (see the Appendix).

In what appears to be a note added in proof to this paper,
Einstein remarked that he had in the meantime convinced himself
that field equations for the theory are not obtained by a
variational principle but by other considerations.

The following paper again appeared in the Academy's {\it
Sitzungberichte} and was presented to the Academy for publication
on January 10. It indeed advanced a new derivation of field
equations that did not make use of a Hamiltonian principle. In the
paper Einstein also introduced a few new notational
conventions.\footnote{The notational idiosyncrasies associated
with Einstein's {\it Fernparallelismus} approach are summarized in
the Appendix.}

The reception of this paper in the public should remind us that
nothing Einstein did at the time took place in an ivory tower.
F\"olsing gives a vivid account of the immense public interest in
Einstein's new theory.\footnote{\cite[pp.~604ff]{Foelsing1997};
see also \cite[p.~346]{Pais1982}.} The January paper itself was
printed and reprinted several times by the Prussian Academy with a
record number of copies. The public interest in Einstein's new
field theory is exemplified by the following quote from a letter
by Eddington who was acknowledging receipt of copies of Einstein's
recent papers, among them \cite{Einstein1929b}:
\begin{quote}
You may be amused to hear that one of our great Department Stores
(Selfridges) has pasted up in its window your paper (the six pages
pasted up side by side) so that passers by can read it all
through. Large crowds gather round to read it!\footnote{Eddington
to Einstein, 11 February, 1929, EA~9-292.}
\end{quote}

The craze apparently had begun with an article in the New York
Times of 4 November 1928 under the title ``Einstein on Verge of
Great Discovery; resents Intrusion.'' The author of this article,
Paul D.\ Miller, gave an account of how he had succeeded to visit
Einstein in his Berlin home. It is a striking example of grooming
the myth of this mysteriously creative genius. The sick Einstein
supposedly ``sat on a sunny beach and appeased his desire to work
by playing his violin to the waves'' but then came up with a new
theory that ``will startle the world far more than relativity
did.'' The article, in any case, seems to have triggered the
interest of numerous other journalists in Einstein's new work.

The journalists, thus alerted of those great events in science,
may have been all too glad to learn that, in early January,
another publication on this new theory appeared and warranted
press coverage. In any case, on January 12, two days after the
submission of \cite{Einstein1929b} to the Academy, the front page
of the New York Times again informed their readers that ``Einstein
Extends Relativity Theory.'' The subtitle: ```Book', Consisting of
Only Five Pages, Took Berlin Scientist Ten Years to Prepare'' may
help to explain why the management of Selfridges came up with the
idea of attracting the curiosity of possible clients by putting up
a copy of this marvel in their window. An English translation of
the note, including all formulas, appeared on the title page of
the {\it New York Herald Tribune} on February 1. And in response
to the overwhelming public interest in his new theory, Einstein
published two popular and non-technical accounts of the latest
developments in the New York Times on February 3
\cite{Einstein1929c} and in the London Times of February 4
\cite{Einstein1929d}.

The essays are a {\it tour-de-force} through the history of field
theory. At its very end, Einstein gave a characterization of
distant parallelism by illustrating the effect of torsion. He has
the reader consider two parallel lines $E_1L_1$ and $E_2L_2$ and
on each a point $P_1$, resp.\ $P_2$. On the first line, $E_1L_1$,
one now chooses another point $Q_1$. Torsion is then expressed by
the fact that parallelograms do not close.
\begin{quote}
If we now draw through $Q_1$ a straight line $Q_1-R$ parallel to
the straight line $P_1$, $P_2$, then in Euclidean geometry this
will cut the straight line $E_2L_2$; in the geometry now used the
line $Q_1-R$ and the line $E_2L_2$ do not in general cut one
another. \cite{Einstein1929c}
\end{quote}
Einstein added
\begin{quote}
To this extent the geometry now used is not only a specialization
of the Riemannian but also a generalization of the Euclidean
geometry. (ibid.)
\end{quote}
In the final paragraph he then stated the expectation that the
solution to the mathematical problem of the correct field laws
would be given by ``the simplest and most natural conditions to
which a continuum of this kind can be subjected.'' Einstein
concluded that
\begin{quote}
the answer to this question which I have attempted to give in a
new paper yields unitary field laws for gravitation and
electromagnetism. (ibid.)
\end{quote}

The unspecific title of the January paper in the {\it
Sitzungsberichte} (``On the Unified Field Theory'') may have
helped to deceive the public about the real content of this rather
specific and technical communication. The title of the next paper
on the {\it Fernparallelismus} approach would surely have been
less attractive for a general public. It is entitled ``Unified
Field Theory and Hamiltonian Principle'' \cite{Einstein1929e}. It
addressed an objection raised by Lanczos and M\"untz. They had
objected that the compatibility of the field equations of the
previous note was not established by the failure to identify four
identical relations between them. Einstein now returned to the
variational approach and gave a Hamiltonian formulation of the
field equation which thus would also guarantee their
compatibility.

\subsection{The field equations}
\label{sec:adolescence2}

Let us know take a closer look at the problem of finding and
justifying field equations within the teleparallel framework. The
tetrad field $h_{s\mu}$ defines both the metric tensor field
$g_{\mu\nu}$, see eq.~(\ref{eq:gintermsofh}), and the
electromagnetic vector potential $\phi_{\mu}$, see
eq.~(\ref{eq:phiintermsofh}). Its sixteen components are the
dynamical variables of the theory. The fundamental question
therefore arises as to the field equations that determine the
tetrad field. Einstein had first discussed this question in his
second note of June 14, 1928, but doubts were raised in the sequel
about the correct field equations and their derivation. These
doubts remained alive with Einstein until the very end of the {\it
Fernparallelismus} episode and are also the major reason for
eventually giving up the teleparallel approach.

We will here review the early attempts at finding field equations
and their derivations as put forward by Einstein in the course of
elaborating the implications of distant parallelism. A closer
analysis of the chronology reveals that Einstein wavered between
two distinct approaches to find, derive, and justify field
equations. Along one approach, he was starting from a variational
principle and was looking for the correct Lagrangian. Along
another approach he was trying to find a set of overdetermined
field equations plus a number of mathematical identities.

The existence of two distinct approaches is strongly reminiscent
of the heuristics followed for the {\it Entwurf} theory, see
sec.~\ref{sec:conclusion} below. And as was the case with the
reconstruction of the genesis and demise of the {\it Entwurf}, the
dynamics of going from one approach to the other, it seems to me,
can only be reconstructed with some confidence on the basis of
more information taken from contemporary correspondence and
research manuscripts. The following sketch will therefore
necessarily have a preliminary character.

\subsubsection{The variational approach}
\label{sec:adolescence2a}

The field equations advanced in Einstein's second note on the
distant parallelism approach were defined by demanding that the
variation of a scalar and globally Lorentz-invariant action
integral $\int \mathcal{H}d\tau$ with respect to the components of
the tetrad field $h_{a\mu}$ vanish, see eq.~(\ref{eq:varpinciple})
above. The Lagrangian ${\mathcal H}$ entering the action integral
had been given in terms of the invariant
$g^{\mu\nu}\Lambda^{\alpha}_{\mu\beta}\Lambda^{\beta}_{\nu\alpha}$
as in eq.~(\ref{eq:Rx175L}).

Einstein did not give any motivation for this kind of Lagrangian.
But it would be a natural {\it ansatz} for him to try. The torsion
tensor $\Lambda^{\alpha}_{\mu\nu}$ was the crucial new quantity of
the theory and the invariant was the simplest combination that was
invariant both for general coordinate transformations and for
rotation of the tetrads. But the torsion tensor allows for
different ways to form a scalar expression. Let us recall then
that in a little note added in proof, Einstein already observed
that ``similar results are obtained'' on the basis of the
Lagrangian
${\mathcal H} = h g_{\mu\nu}g^{\alpha\sigma}g^{\beta\tau}
\Lambda^{\mu}_{\alpha\beta}\Lambda^{\nu}_{\sigma\tau}$, see
(\ref{eq:Rx175Ladd}) above. Einstein commented that ``for the time
being'' there was an uncertainty regarding the choice of
${\mathcal H}$. It is unclear whether there was an external
trigger for this realization.

But things got worse. In his contribution to the
Stodola-Festschrift Einstein briefly sketched his new approach.
With respect to the derivation of field equations, Einstein now
considered a generic Lagrangian
\begin{equation}
{\mathcal H} = h (A J_1 + B J_2 + C J_3)
\label{eq:StodolaLgeneric}
\end{equation}
where
\begin{align}
J_1 &=
g^{\mu\nu}\Lambda^{\alpha}_{\mu\beta}\Lambda^{\beta}_{\nu\alpha},
\label{eq:Winvariant1} \\
J_2 &=
g^{\mu\nu}\Lambda^{\alpha}_{\mu\alpha}\Lambda^{\beta}_{\nu\beta},
\label{eq:Winvariant2} \\
J_3 &= g^{\mu\sigma}g^{\nu\tau}g_{\lambda\rho}
\Lambda^{\lambda}_{\mu\nu}\Lambda^{\rho}_{\sigma\tau}.
\label{eq:Winvariant3}
\end{align}
Although Einstein does not explicitly refer to Weitzenb\"ock's
paper \cite{Weitzenboeck1928} in the Stodola-Festschrift, it
should be pointed out that the three terms $J_i$ are explicitly
listed in that paper (p.~470) as the only invariants (under both
general coordinate transformations and rotations of the tetrads)
of second degree in $\Lambda^{\nu}_{\alpha\beta}$.

Einstein remarked that
\begin{quote}
The elaboration and physical interpretation of the theory is made
difficult by the lack of an apriori constraint for choosing the
ratio of the constants $A$, $B$, $C$.\footnote{``Die Ausarbeitung
und physikalische Interpretation der Theorie wird dadurch
erschwert, dass f\"ur die Wahl des Verh\"altnisses der Konstanten
$A$, $B$, $C$ ap priori keine Bindung vorhanden ist.''
\cite[p.~132]{Einstein1929a}.}
\end{quote}

Obviously, his first ansatz eq.~(\ref{eq:Rx175L}) of the June 14
note is contained in the generic Lagrangian
(\ref{eq:StodolaLgeneric}) by specifying to $A=1$, $B=C=0$. The
alternative Lagrangian (\ref{eq:Rx175Ladd}) advanced in the note
added in proof to that note would be given by specifying to the
case of $C=1$, $A=B=0$. In the Stodola-Festschrift, Einstein then
specified to the case $B=-A$, $C=0$, which would read explicitly
\begin{equation}
{\mathcal H} = h
\bigg(g^{\mu\nu}\Lambda^{\alpha}_{\mu\alpha}\Lambda^{\beta}_{\nu\beta}
-
g^{\mu\nu}\Lambda^{\alpha}_{\mu\beta}\Lambda^{\beta}_{\nu\alpha}\bigg).
\end{equation}
He observed, however, that the specialization $B=-A$, $C=0$ should
be taken only at the level of the field equations, not on the
level of the variational principle. Otherwise, the electromagnetic
field equations would not be obtained. This arcane remark is not
further explained by explicit calculations.

Moreover, in what appears to be a note added in proof to the
Stodola-{\it Festschrift}, Einstein remarked that he had in the
meantime convinced himself that the ``most natural'' {\it
ans\"atze} for the field equations are not obtained on the basis
of a Hamiltonian principle.\footnote{``Inzwischen hat mich eine
tiefere Analyse der allgemeinen Eigenschaften der Strukturen der
oben entwickelten Art zu der \"Uberzeugung gef\"uhrt, dass die
nat\"urlichsten Ans\"atze f\"ur die Feldgleichungen nicht aus
einem Hamilton-Prinzip, sondern auf anderem Wege zu gewinnen sind.
\cite[p.~132]{Einstein1929a}.} For a different approach, he
referred to his new paper ``On the Unified Field Theory'' in the
Prussian Academy proceedings \cite{Einstein1929b}.

However, that alternative approach of January 10, which will be
discussed below, was shortlived. Already some two months later, on
21 March 1929, Einstein returned to the variational approach for
deriving the field equations.

Since Einstein had introduced in the January 10 paper a number of
new conventions, the notation used in the March note is slightly
different from the notation used in the Stodola-Festschrift. Thus,
he had dropped a factor of $1/2$ in the definition of the torsion
and he had introduced an idiosyncratic convention of indicating
raising and lowering indices by underlining them. He also used a
slightly different notation for the terms defined in
eqs.~(\ref{eq:Winvariant1}), (\ref{eq:Winvariant2}),
(\ref{eq:Winvariant3}) using ${\mathcal J} = hJ$, {\it and} he
renumbered two terms, i.e.\ he has ${\mathcal J}_2\equiv hJ_3$
resp.\ ${\mathcal J}_3\equiv hJ_2$. If we keep with the notation
of the Stodola-Festschrift
(\ref{eq:StodolaLgeneric},\ref{eq:Winvariant1}-\ref{eq:Winvariant3})),
Einstein now advanced the following Lagrangian (up to an overall
constant)
\begin{equation}
{\mathcal H} = h\bigg( \frac{1}{2}J_1 - J_2 + \frac{1}{4}J_3
\bigg). \label{eq:HamPrinL}
\end{equation}
This Lagrangian is explicitly justified by the following two
postulates. ${\mathcal H}$ must be a function of second degree in
the torsion tensor $\Lambda^{\alpha}_{\mu\nu}$ which makes it a
linear combination of the three terms $J_1$, $J_2$,
$J_3$.\footnote{This fact is stated in
\cite[p.~470]{Weitzenboeck1928}.} Second, the resulting field
equation must be symmetric in the free indices, and Einstein
claimed that this postulate uniquely fixes the specific linear
combination (\ref{eq:HamPrinL}).

More specifically, Einstein claimed that the combination
(\ref{eq:HamPrinL}) produces only one part of the field equations,
i.e.\ the part that reduces to the gravitational field equation in
linear approximation. In order to obtain the electromagnetic field
equations, he proposed to consider a slightly distorted Lagrangian
\begin{equation}
\bar{\mathcal{H}} = \mathcal{H} +
h\epsilon_1\left(\frac{1}{2}J_1-\frac{1}{4}J_2\right) -
h\epsilon_2J_3 \label{eq:Lagrepsilon}
\end{equation}
where the existence of electric charges demands taking the limit
$\epsilon_2/\epsilon_1\rightarrow 0$. In that limit, the relation
\begin{equation}
S^{\alpha}_{\underline{\mu\nu}}=0 \label{eq:S}
\end{equation}
is obtained where the quantity $S^{\alpha}_{\underline{\mu\nu}}$
was defined as the completely antisymmetrized torsion
\begin{equation}
S^{\underline{\alpha}}_{\mu\nu} =
\Lambda^{\underline{\alpha}}_{\mu\nu} +
\Lambda^{\underline{\nu}}_{\alpha\mu} +
\Lambda^{\underline{\mu}}_{\nu\alpha}, \label{eq:torsionantisym}
\end{equation}
using Einstein's temporary convention to indicate a raising resp.\
lowering of an index by underlining, see (\ref{eq:raiseindex})
below. Einstein claimed that the relation (\ref{eq:S}) implies
that the combination (\ref{eq:HamPrinL}) is equivalent to the
earlier combination $J_1-J_2$ of \cite{Einstein1929a}.

The procedure of varying a slightly distorted Lagrangian in order
to obtain the electromagnetic field equation had been developed
partly within the overdetermination approach. The details were
not, however, spelled out explicitly in the published papers on
the {\it Fernparallelismus} approach.

In summary, Einstein had advanced four different field equations
in three papers which are given by the generic Lagrangian
(\ref{eq:StodolaLgeneric}) and the coefficients
\begin{center}
\begin{tabular}{|l|l|c|c|c|}
\hline Paper & Date & $A$ & $B$ & $C$ \\
\hline
\cite{Einstein1928b} & 14 Jun 28 & 1 & 0 & 0 \\
\cite[note added]{Einstein1928b} & after 14 Jun 28 & 0 & 0 & 1 \\
\cite{Einstein1929a} & 10 Dec 28 & 1 & - 1 & 0 \\
\cite{Einstein1929b} & 21 Mar 29 & $1/2$ & $-1$
& $1/4$ \\
\hline
\end{tabular}
\end{center}

We shall now turn to the second approach of deriving field
equations for the teleparallel theory.

\subsection{The overdetermination approach}
\label{sec:adolescence2b}

Already by the end of 1928, around the time when he had submitted
his paper for the Stodola-Festschrift, Einstein may have become
dissatisfied with the variational approach. This may have been due
mainly to the fact that he did not succeed in finding a convincing
way of getting unique field equations. But there were also other
difficulties associated with the demands that the electromagnetic
field equations should be obtained in the linearized approximation
and that non-singular, spherically symmetric and stationary,
charged or massive solutions to the field equations should exist.

In any case, a few days after sending off his manuscript for the
Stodola Festschrift, he wrote to Hermann M\"untz
\begin{quote}
I have had a simple, cheeky idea which throws the Hamiltonian
principle overboard. The cart shall now be put before the horse: I
choose the field equations in such a way that I am sure that they
imply the Maxwellian equations.\footnote{``Ich habe eine einfache,
freche Idee gehabt, die das Hamilton'sche Prinzip \"uber Bord
wirft. Das Pferd soll nun vom Schwanze aus aufgez\"aumt werden:
ich w\"ahle die Feldgleichungen so, dass ich sicher bin, dass sie
die Maxwellschen Gleichungen zur Folge haben.'' Einstein to
M\"untz, 13 December 1928, EA 18-317.}
\end{quote}
The idea was to use an identity which implies the validity of the
Maxwell equations and construct field equations by the demand that
this identity was automatically satisfied. But again Einstein
encountered technical difficulties showing him that the simple
idea was not feasible.
\begin{quote}
The derivation of the field equations by means of the identity is
a task that is more subtle than I originally
thought.\footnote{``Die Aufstellung der Feldgleichungen mit Hilfe
der Identit\"at ist eine subtilere Aufgabe, als ich urspr\"unglich
dachte.'' Einstein to M\"untz, 15 December 1928, EA~18-318.}
\end{quote}
However, he did pursue the general approach further and soon came
up with another derivation of the field equations.

The next paper then was the January note that would attract so
much public interest \cite{Einstein1929b}. It presented a
different approach to a derivation of field equation since that
derivation on the basis of a Hamiltonian principle had not ``led
to a simple and completely unique path.''\footnote{``[...]
f\"uhrte die Ableitung der Feldgleichung aus dem Hamiltonschen
Prinzip auf keinen einfachen und v\"ollig eindeutigen Weg. Diese
Schwierigkeiten verdichteten sich bei genauerer \"Uberlegung.''
\cite[p.~2]{Einstein1929b}.}

Einstein now argued like this. He first derived two sets of
identities for the torsion tensor $\Lambda^{\alpha}_{\mu\nu}$. The
first identity was obtained by starting from the vanishing of the
Riemann curvature (\ref{eq:Riemanncurvature}) for the
Weitzenb\"ock connection (\ref{eq:Weitzconn}). If
(\ref{eq:Riemanncurvature}) is cyclically permuted in the lower
indices and added it produces the identity
\begin{equation}
0\equiv\Lambda^{\iota}_{\kappa\lambda,\mu} +
\Lambda^{\iota}_{\lambda\mu,\kappa} +
\Lambda^{\iota}_{\mu\kappa,\lambda} +
\Delta^{\iota}_{\sigma\kappa}\Lambda^{\sigma}_{\lambda\mu} +
\Delta^{\iota}_{\sigma\lambda}\Lambda^{\sigma}_{\mu\kappa} +
\Delta^{\iota}_{\sigma\mu}\Lambda^{\sigma}_{\kappa\lambda},
\label{eq:identity1}
\end{equation}
which can be rewritten using covariant derivatives (with respect
to the connection $\Delta^{\iota}_{\sigma\kappa}$),
\begin{equation}
\Lambda^{\iota}_{\kappa\lambda;\mu} =
\Lambda^{\iota}_{\kappa\lambda;\mu} +
\Lambda^{\sigma}_{\kappa\lambda}\Delta^{\iota}_{\sigma\mu} -
\Lambda^{\iota}_{\sigma\lambda}\Delta^{\sigma}_{\kappa\mu},
\end{equation}
as
\begin{equation}
0 \equiv \Lambda^{\iota}_{\kappa\lambda;\mu} +
\Lambda^{\iota}_{\lambda\mu;\kappa} +
\Lambda^{\iota}_{\mu\kappa;\lambda} +
\Lambda^{\iota}_{\kappa\alpha}\Lambda^{\alpha}_{\lambda\mu} +
\Lambda^{\iota}_{\lambda\alpha}\Lambda^{\alpha}_{\mu\kappa} +
\Lambda^{\iota}_{\mu\alpha}\Lambda^{\alpha}_{\kappa\lambda}.
\label{eq:identity2}
\end{equation}
Contraction of (\ref{eq:identity2}) and using $\phi_{\mu}\equiv
\Lambda^{\alpha}_{\mu\alpha}$, see (\ref{eq:phiintermsofh}), the
identity can be written as
\begin{equation}
0 \equiv \Lambda^{\alpha}_{kl;\alpha} + \phi_{l;k} - \phi_{k;l} -
         \phi_{\alpha}\Lambda^{\alpha}_{kl},
\label{eq:identity3}
\end{equation}
and, introducing the tensor density
\begin{equation}
\mathfrak{B}^{\alpha}_{\kappa\lambda} =
h\left(\Lambda^{\alpha}_{\kappa\lambda} +
\phi_{\lambda}\delta^{\alpha}_{\kappa} -
\phi_{\kappa}\delta^{\alpha}_{\lambda}\right), \label{eq:B}
\end{equation}
(\ref{eq:identity3}) can further be rewritten as
\begin{equation}
\mathfrak{B}^{\alpha}_{\kappa\lambda;\alpha} -
\mathfrak{B}^{\sigma}_{\kappa\lambda}\Lambda^{\alpha}_{\sigma\alpha}
\equiv \mathfrak{B}^{\alpha}_{\kappa\lambda/\alpha}\equiv 0,
\label{eq:identity4}
\end{equation}
i.e.\ as the vanishing of some special divergence denoted by
$\dots_{/\alpha}$. This notation shall be temporarily used here,
too, in order to have a chance to see Einstein's heuristics in his
line of argument.

The second identity was derived by considering the commutator of
the covariant derivatives for an arbitrary tensor
$T^{\dots}_{\dots}$,
\begin{equation}
T^{\dots}_{\dots;\iota;\kappa}-T^{\dots}_{\dots;\kappa;\iota} =
-T^{\dots}_{\dots;\sigma}\Lambda^{\sigma}_{\iota\kappa}.
\label{eq:commutationlaw}
\end{equation}
Inserting $\mathfrak{B}^{\sigma}_{\kappa\lambda}$ for
$T^{\dots}_{\dots}$, rewriting in terms of $\dots_{/\alpha}$, and
using the first identity (\ref{eq:identity4}), Einstein obtained
the second identity as
\begin{equation}
\left(\mathfrak{B}^{\alpha}_{\underline{\kappa}\underline{\lambda}/\lambda}
- \mathfrak{B}^{\sigma}_{\underline{\kappa}\underline{\tau}}
\Lambda^{\alpha}_{\sigma\tau}\right)_{/\alpha} = 0.
\label{eq:identity5}
\end{equation}

Field equations are now derived as follows. Identity
(\ref{eq:identity4}) motivated Einstein to consider the vanishing
of ``the other divergence,'' i.e.\
\begin{equation}
\mathfrak{B}^{\alpha}_{\underline{\kappa}\underline{\lambda}/\lambda}
= 0,
\end{equation}
as the field equation. In linear approximation, he obtained indeed
the gravitational equations  but could not get the electromagnetic
equations, a difficulty that he traced back to the identity
\begin{equation}
\mathfrak{B}^{\alpha}_{\underline{\kappa}\underline{\lambda}/\lambda/\alpha}
=
\mathfrak{B}^{\alpha}_{\underline{\kappa}\underline{\lambda}/\alpha/\lambda}.
\end{equation}
The trick to get also the electromagnetic equations was to look at
the quantity
\begin{equation}
\overline{\mathfrak{B}}^{\alpha}_{\kappa\lambda} =
\mathfrak{B}^{\alpha}_{\kappa\lambda} + \epsilon
h\left(\phi_{\lambda}\delta^{\alpha}_{\kappa} -
\phi_{\kappa}\delta^{\alpha}_{\lambda}\right)
\end{equation}
and take as field equation
\begin{equation}
\overline{\mathfrak{B}}^{\alpha}_{\underline{\kappa}\underline{\lambda}/\lambda}
= 0.
\end{equation}
Maxwell's equations would then be obtained by taking the
divergence with respect to the index $\alpha$. The gravitational
equations would still be obtained by taking the limit of
$\epsilon\rightarrow 0$.

Going beyond the linear approximation, Einstein now started from
the identity (\ref{eq:identity5}), and postulated the field
equations
\begin{equation}
\overline{\mathfrak{B}}^{\alpha}_{\underline{\kappa}\;\underline{\lambda}/\lambda}
-
\overline{\mathfrak{B}}^{\sigma}_{\underline{\kappa}\;\underline{\tau}}
\Lambda^{\alpha}_{\sigma\tau} = 0,
\end{equation}
where again the electromagnetic equations are obtained by
considering the divergence with respect to the index $\alpha$ and
the gravitational equations by taking the limit
$\epsilon\rightarrow 0$. Consequently, the final field equations
are
\begin{equation}
\mathfrak{B}^{\alpha}_{\underline{\kappa}\;\underline{\lambda}/\lambda}
-
\mathfrak{B}^{\sigma}_{\underline{k}\;\underline{\tau}}\Lambda^{\alpha}_{\sigma\tau},
\label{eq:fieldeqRx183a}
\end{equation}
and
\begin{equation}
[h \left(\phi_{\underline{k};\underline{\alpha}} -
         \phi_{\underline{\alpha};\underline{k}}\right)]_{/\alpha} =
         0.
\label{eq:fieldeqRx183b}
\end{equation}
These are $20$ equations for the sixteen quantities $h_{a\mu}$.
The compact notation involves the idiosyncratic notation of the
divergence $\dots_{/\alpha}$ introduced in (\ref{eq:identity4}),
the convention of raising indices by underlining them according to
(\ref{eq:raiseindex}), and the introduction of the quantities
$\mathfrak{B}^{\alpha}_{\kappa\lambda}$ in (\ref{eq:B}),
$\Lambda^{\alpha}_{\mu\nu}$ in (\ref{eq:Lambda}), and
$\phi_{\alpha}$ in (\ref{eq:phiintermsofh}). Einstein argued that
there were $8$ identities between these $20$ equation. But he had
explicitly given only four of them, i.e.\ (\ref{eq:identity5}).
The problem here was that Einstein had erroneously assumed the
existence of a set of identities compatible with the field
equations, as pointed out to him soon by Lanczos and M\"untz.

\section{The {\it Mathematische Annalen} paper}
\label{sec:maturity}

The overdetermination approach had produced field equations
(\ref{eq:fieldeqRx183a}) and (\ref{eq:fieldeqRx183b}) and the
variational approach had produced Lagrangian
(\ref{eq:Lagrepsilon}). It is unclear to me to what extent
Einstein reflected on the compatibility of the two approaches,
i.e.\ to what extent he tried to produce the same set of field
equations along the two approaches, or specifically how the
Lagrangian (\ref{eq:Lagrepsilon}) published in March relates to
the field equations (\ref{eq:fieldeqRx183a}) and
(\ref{eq:fieldeqRx183b}) of January. In any case, it should have
become clear that all explicit calculations in terms of the
fundamental tetrad variables $h_{a\mu}$ involved an appreciable
amount of algebraic complexity, and it seems that many
implications were only realized on the level of the linear
approximation.

The theory of distant parallelism reached its mature stage in the
summer of 1930 with Einstein's major publication concerning the
{\it Fernparallelismus} approach: a review paper that was
published in the {\it Mathematische Annalen} \cite{Einstein1930a}.
The publication history of this paper is a little involved and
reflects an issue of priority that arose between Einstein and Elie
Cartan. The paper also gave a new derivation of the final field
equations along the overdetermination approach.

\subsection{The publication history}
\label{sec:maturity1}

The prehistory of this paper seems to begin with a letter by Elie
Cartan that was sent to Einstein on 8 May 1929 and that triggered
an extensive correspondence between the two
scientists.\footnote{This correspondence was published in
\cite{Debever1979}.} In this first letter, Cartan pointed out to
Einstein that the mathematical framework of Einstein's {\it
Fernparallelismus} was, indeed, a special case of a generalization
of Riemannian geometry advanced by him in previous years
\begin{quote}
Now, the notion of Riemannian space endowed with a {\it
Fernparallelismus} is a special case of a more general notion,
that of space with a Euclidean connection, which I outlined
briefly in 1922 in an article in the {\it Comptes Rendues}
[...]\footnote{``Or la notion d'espace riemannien dou\'e d'un {\it
Fernparallelismus} est un cas particulier d'une notion plus
g\'en\'erale, celle d'espace \'a connexion euclidienne, que j'ai
indiqu\'ee succinctement en 1922 dans une note des {\it Comptes
rendus}.'' Cartan to Einstein, 8 May 1929,
\cite[Doc.~I]{Debever1979}.}
\end{quote}
The reference is to \cite{Cartan1922} and what Cartan here calls a
Euclidean connection is a non-symmetric linear connection on a
real, differentiable manifold, thus allowing for both Riemannian
curvature and torsion.\footnote{See
\cite[ch.~7]{AkivisRosenfeld1993} for an account of Cartan's work
on generalized spaces.} Moreover, Cartan pointed out that he had
even spoken to Einstein about this generalized geometry when they
had met, in 1922, at Hadamard's home. He even remembered that he
had tried to illustrate the case of teleparallelism in his theory
to Einstein on this occasion.

On receiving this letter, Einstein seems to have been quick to
react. Apparently he sent off a review article of his theory to
the {\it Zeitschrift der Physik} on the next day. This review
article never appeared. In fact, it may have been sent off at the
time prematurely only because Einstein, in his response to Cartan,
another day later, wanted to mention this work of his. What he
wrote to Cartan essentially was an acknowledgment that Cartan was
right:
\begin{quote}
I see, indeed, that the manifolds used by me are a special case
studied by you.\footnote{``Ich sehe in der That ein, dass die von
mir benutzten Mannigfaltigkeiten in den von Ihnen studierten als
Spezialfall enthalten sind.'' Einstein to Cartan, 10 May 1929,
\cite[Doc.~II]{Debever1979}.}
\end{quote}
By way of excuse, he pointed out that Weitzenb\"ock had already
written a review article on the mathematical foundations of
teleparallelism with a supposedly complete bibliography but had
failed to cite Cartan's work. And in his own review article of the
previous day, he himself, so he wrote, had not mentioned any
literature at all, not even his own papers.

But Einstein acknowledged Cartan's claim of priority and suggested
that Cartan write a brief historical account, ``a short analysis
of the mathematical background,'' to be appended to his own paper
but under Cartan's name.\footnote{``Schreiben sie \"uber diese
mathematische Vorgeschichte eine kurze Charakteristik, die wir
meiner neuen zusammenfassenden Arbeit anheften, nat\"urlich unter
Ihrem Namen, aber mit meiner Arbeit zu einem Ganzen vereinigt.''
ibid. One cannot help to be reminded of Einstein's and Grossmann's
earlier {\it Outline of a Generalized Theory of Relativity and of
a Theory of Gravitation}. But in that case, Einstein and Grossmann
had actually collaborated to obtain the results presented in their
paper.}

Cartan agreed in a letter of 15 May and, indeed, sent a manuscript
to Einstein a little more than a week later, i.e.\ on May 24th.

One would think that Einstein, on receiving Cartan's manuscript
would have forwarded it to the {\it Zeitschrift f\"ur Physik} as
he had suggested to Cartan. It seems, however, that Einstein did
not do so. What might have changed matters was perhaps a letter by
Lanczos that Einstein may have received on the very same day,
since the latter had written it the day before. In his
correspondence, Lanczos communicated to Einstein his insight that
in Weitzenb\"ock's theory the scalar Riemannian curvature $R$ is
essentially equivalent to ``the invariant preferred by
you,''\footnote{``die von Ihnen bevorzugte Invariante,'' Lanczos
to Einstein, 23 May 1929, EA~15-230.} $\frac{1}{2}J_1 +
\frac{1}{4}J_2 - J_3$, plus a divergence. From this result, it
clearly follows that the variational principle would not allow to
derive the electromagnetic equations.

In a letter to M\"untz, written a few days later, Einstein wrote:
\begin{quote}
Regarding the whole problem Lanczos' discovery changes the
situation profoundly.\footnote{``Was das ganze Problem anlangt, so
\"andert Lanczos Entdeckung die Situation grundlegend.'' Einstein
to M\"untz, 27 May 1929, EA 18-323. In other words, what Lanczos
had seen was that distant parallelism is essentially equivalent to
classical general relativity, a fact that was fully realized only
much later.}
\end{quote}
No paper by Einstein or Cartan appeared in the {\it Zeitschrift
f\"ur Physik}. Nevertheless, the next paper on teleparallelism by
Einstein was a review paper and its aim was to present the theory
in a self-contained way without reference to earlier publications.
This work appeared in the {\it Mathematische Annalen} with a
historical review paper on the subject by Cartan appended to it in
the very same issue of this journal.\footnote{This is another
similarity to the Einstein-Grossmann collaboration. Both papers
that were coauthored by Einstein and Grossmann appeared in the
more mathematically oriented {\it Zeitschrift f\"ur Mathematik und
Physik}, whereas Einstein published most of his other notes at the
time in the {\it Annalen der Physik}.}

Einstein had been co-editor of the {\it Mathematische Annalen}
from 1919 until 1928.\footnote{For a historical account of the
changes in the editorial board of the {\it Mathematischen Annalen}
in this period, see \cite{vanDalen1990}.} However, during that
time he had published only a single paper in this journal himself
\cite{Einstein1927}. The paper on teleparallelism would be his
only other paper published in the {\it Mathematische Annalen}.

Einstein's paper in the {\it Annalen} is entitled `` Unified field
theory based on the Riemann metric and on distant parallelism''
{\cite{Einstein1930a}. According to the published version it was
received by the {\it Annalen} on 19 August 1929. But in a letter
to the managing editor Otto Blumenthal, dated 19 August 1929,
Einstein only announced submission of ``an already completed
summarizing work on the mathematical apparatus of the general
field theory''\footnote{``eine bereits fertiggestellte
zusammenfassende Arbeit \"uber den mathematischen Apparat der
allgemeinen Feldtheorie,'' Einstein to Blumenthal, 19 August 1929,
EA~9-005.} to the {\it Annalen}. In the letter, Einstein required
whether ``a treatise in the French language (ca.\ 12 pages long)
on the prehistory of the problem''\footnote{``eine franz\"osisch
geschriebene Abhandlung (etwa 12 Seiten L\"ange) \"uber die
Vorgeschichte des Problems'' ibid.} composed by Cartan could be
appended to his paper. He also enquired how long it would take
until the paper would be printed.

A week later Einstein informed Cartan of the change regarding his
publication plans and apologized for the long silence which was
\begin{quote}
caused by many doubts as to the correctness of the course I have
adopted. But now I have come to the point that I am persuaded I
have found the simplest legitimate characterization of a
Riemannian metric with distant parallelism that can occur in
physics.\footnote{``verursacht durch viele Zweifel an der
Richtigkeit des eingeschlagenen Weges. Nun aber bin ich soweit
gekommen, dass ich die einfachste gesetzliche Charakterisierung
einer Riemann-Metrik mit Fernparallelismus, welche f\"ur die
Physik in Betracht kommen kann, gefunden zu haben \"uberzeugt
bin.'' Einstein to Cartan, 25 August 1929,
\cite[Doc.~V]{Debever1979}.}
\end{quote}
Einstein added that he now wanted to publish in the {\it Annalen}
since ``for the time being only the mathematical implications are
explored and not their application to
physics.''\footnote{``einstweilen nur die mathematischen
Zusammenh\"ange untersucht werden, nicht aber deren Anwendung auf
die Physik.'' ibid.}

Blumenthal only responded on September 9 to Einstein's enquiry,
agreeing to the proposal and informing Einstein that the
publication will be delayed by approximately six months. A few
days later, on 13 September, Einstein finally sends both
manuscripts, his own and Cartan's, to Blumenthal for publication
in the {\it Annalen}. In the covering letter, he expressed his
understanding for the delay in publication but added
\begin{quote}
However, it is a pity because it delays the collaboration of the
colleagues on this problem which is fundamental and, after the
most recent results, really promising. Physics after all has a
different rhythm than mathematics.\footnote{``Es ist aber schade,
weil die Mitarbeit der Kollegen an diesem fundamtentalen und nach
den letzten Ergebnissen wirklich aussichtsreichem Problem dadurch
verz\"ogert wird. Die Physik hat eben einen anderen Rhytmus als
die Mathematik.'' Einstein to Blumenthal, 13 September 1929,
EA~9-009.}
\end{quote}
Proofs of the paper were probably received by late
November.\footnote{In a letter to Einstein, dated 3 December,
Cartan informed him that he had already returned the proofs which
he had received ``a few days ago.'' \cite[Doc.~VII]{Debever1979}.}
According to the title page of the pertinent issue of the {\it
Annalen}, it was ``completed'' (``abgeschlossen'') on 18 December
1929.

\subsection{The derivation of the field equations}
\label{sec:maturity2}

Einstein's {\it Annalen} paper has five paragraphs. It begins with
an exposition of the mathematical structure of Fernparallelismus
in the first three paragraphs. Here he reverted to the original
notation of writing both indices of the tetrads to the right,
using latin character for algebra, greek character for coordinate
indices. He also explicitly commented that he would no longer use
the new divergence operation. In paragraphs four and five, he then
discussed the field equations and their first approximation.

As pointed our explicitly in the introductory paragraph of the
paper, the most important and in any case new part of the paper
concerns the ``derivation of the simplest field laws to which a
Riemannian manifold with teleparallelism may be
subjected.''\footnote{``die Auffindung der einfachsten
Feldgesetze, welchen eine Riemannsche Mannigfaltigkeit mit
Fern-Parallelismus unterworfen werden kann.''
\cite[p.~685]{Einstein1930a}.} Here, however, he no longer
proceeded along a variational approach but argued like follows.

Einstein observed that the simplest field equations that one is
looking for would be conditions on the torsion tensor
$\Lambda^{\;\mu}_{\alpha\;\nu}$ expressed in terms of the
Weitzenb\"ock connection $\Delta^{\;\mu}_{\alpha\;\nu}$ resp.\ in
terms of the the tetrad fields $h_{a\mu}$ as in (\ref{eq:Lambda}).
Although, he does not say so explicitly in the paper, the
rationale for this argument would be that for vanishing torsion
one also has vanishing Riemannian curvature for the Levi-Civita
connection and hence no gravitational field.

He now argues for a heuristics of finding field equations along
the overdetermination approach. Since the tetrad field has $n^2$
components of which $n$ need be undefined due to general
covariance, one needs $n^2-n$ independent field equations. The
heuristic principle of overdetermination is then stated like this:
\begin{quote}
On the other hand it is clear that a theory is all the more
satisfying the more it restricts the possibilities (without
getting into conflict with experience). The number $Z$ of field
equations hence shall be as large as possible. If $\overline{Z}$
is the number of identities between them, then $Z-\overline{Z}$
must be equal to $n^2-n$.\footnote{``Andererseits ist klar,
da{\ss} eine Theorie desto befriedigender ist, je mehr sie die
M\"oglichkeiten einschr\"ankt (ohne mit Erfahrungen in Widerspruch
zu treten). Die Zahl $Z$ der Feldgleichungen soll also m\'oglichst
gro{\ss} sein. Ist $\overline{Z}$ die Zahl der zwischen diesen
bestehenden Identit\"aten, so mu{\ss} $Z-\overline{Z}$ gleich
$n^2-n$ sein.'' \cite[p.~692]{Einstein1930a}.}
\end{quote}

The identity that Einstein now put at the center of his derivation
of field equations is related to identity (\ref{eq:identity5})
since it is similarly obtained using the commutation law
(\ref{eq:commutationlaw}) for covariant differentiation. But now
he no longer used the quantity
$\mathcal{B}^{\alpha}_{\kappa\lambda}$ nor the new divergence
notation $\dots_{/\alpha}$ but rather looked at the commutation of
the covariant derivatives for the torsion tensor
$\Lambda^{\;\mu}_{\alpha\;\nu}$ directly. This produced the
identity
\begin{equation}
\Lambda^{\;\alpha}_{\underline{\mu}\;\underline{\nu};\nu\alpha} -
\Lambda^{\;\alpha}_{\underline{\mu}\;\underline{\nu};\alpha\nu} -
\Lambda^{\;\sigma}_{\underline{\mu}\;\underline{\tau};\alpha}
\Lambda^{\;\alpha}_{\sigma\;\tau} \equiv 0, \label{eq:Id1}
\end{equation}
where again the raising or lowering indices is indicated by
underlining. Introducing the quantities
\begin{align}
G^{\mu\alpha} &\equiv \Lambda^{\;\alpha}_{\underline{\mu}\;\underline{\nu};\nu}
- \Lambda^{\;\sigma}_{\underline{\mu}\;\underline{\tau}}
\Lambda{\;\alpha}_{\sigma\;\tau}, \\
F^{\mu\nu} &\equiv \Lambda^{\;\alpha}_{\underline{\mu}\;\underline{\nu};\alpha}
\end{align}
the identity (\ref{eq:Id1}) can be rewritten as
\begin{equation}
G^{\mu\alpha}_{\quad;\alpha} - F^{\mu\nu}_{\quad;\nu} +
\Lambda^{\;\sigma}_{\underline{\mu}\;\underline{\tau}}F_{\sigma\tau}
= 0. \label{eq:Id2}
\end{equation}

The field equations are now introduced as
\begin{align}
G^{\mu\alpha} &= 0, \label{eq:fieldeq1}\\
F^{\mu\alpha} &= 0. \label{eq:fieldeq2}
\end{align}
As it stands the system of field equations does not satisfy
Einstein's heuristic of overdetermination. Since $F^{\mu\nu}$ is
antisymmetric, equations (\ref{eq:fieldeq1}, \ref{eq:fieldeq2})
represent $n^2 + n(n-1)/2$ field equation which obey only $n$
identities (\ref{eq:Id1}). In order to balance the number of
equations and identities Einstein proceeded to introduce an
equivalent system of $n^2+n$ field equations. Rewriting identity
(\ref{eq:identity3}) as
\begin{equation}
\Lambda^{\alpha}_{\kappa\lambda;\alpha} \equiv
\phi_{\kappa,\lambda} - \phi_{\lambda,\kappa},
\label{eq:identity6}
\end{equation}
he observed that (\ref{eq:fieldeq2}) implies that $\phi_{\alpha}$
may be obtained from a scalar potential $\psi$. Hence,
(\ref{eq:fieldeq2}) together with (\ref{eq:identity6}) is
equivalent to
\begin{equation}
\phi_{\alpha} = \frac{\partial\lg\psi}{\partial x^{\alpha}},
\end{equation}
which increases the number of variables to $n^2+1$ but reduces the
number of field equations to $n^2-n$. Einstein still needed
another identity which he derived by looking at the antisymmetric
part $\underline{G}^{\mu\alpha}$ of $G^{\mu\alpha}$. He obtained a
set of $n$ identities,
\begin{equation}
\left[h\psi\left(2\underline{G}^{\mu\alpha}-F^{\mu\alpha} +
S^{\sigma}_{\underline{\mu}\underline{\alpha}}(\phi_{\sigma} -
(\lg\psi)_{,\sigma})\right)\right]_{\alpha} \equiv 0,
\label{eq:identity7}
\end{equation}
where $S^{\sigma}_{\mu\alpha}$ is the completely antisymmetrized
torsion (\ref{eq:torsionantisym}). Of the $n$ equations
(\ref{eq:identity7}) only $n-1$ are independent since the
antisymmetry of $[\dots]$ with respect to $\alpha$ and $\mu$
implied $[\dots]_{,\alpha\mu}=0$, irrespective of any specific
choice for $G^{\mu\alpha}$ or $F^{\mu\alpha}$. Computing again the
balance of the number of field equations ($n^2+n$) minus the
number of (independent) identities ($n+n-1$) compared to the
number of field variables ($n^2+1$) minus the number of space-time
dimensions to allow for general covariance $n$, these numbers now
added up correctly as
\begin{equation}
(n^2+n) - (n+n-1) = (n^2+1) - n.
\end{equation}
This essentially completed the derivation of the field equations
(\ref{eq:fieldeq1}, \ref{eq:fieldeq2}) as given in the {\it
Annalen} paper. Actually Einstein was a bit more precise by
arguing for the compatibility of the field equations on a
hypersurface $x^n=a$ and the possibility of a smooth continuation
of all relations off the hypersurface. A variational principle is
no longer mentioned. In the final paragraph, Einstein looked at
the first approximation of the field equations and derived
relations that correspond to the Poisson equation and to the
vacuum Maxwell equations, respectively.\footnote{This part of the
{\it Annalen} paper is the subject of a later correspondence that
took place in the late thirties between Einstein and Herbert
E.~Salzer who wrote a master's thesis on ``analytic, geometric and
physical aspects of distant parallelism''. In this correspondence,
Einstein admitted an error in the last section of his {\it
Annalen} paper. But at that time, he had abandoned the approach
long ago anyway. See \cite{Salzer1974} for a detailed discussion.}

\section{The final fate of the approach}
\label{sec:oldage}

The theory had now reached a stage where Einstein essentially
stopped looking for other acceptable field equations, just as in
the case of the prehistory of general relativity with publication
of the {\it Entwurf}. And just as with the {\it Entwurf}, the {\it
Annalen} paper represents the culmination of the distant
parallelism approach. At this point Einstein, accepted the
equations that he had come up with and proceeded to look at their
physical and mathematical consequences. This latter endeavour
involved the elaboration of implications of physical significance
such as the existence of particle-like solutions and their
equations of motion. It also involved, again in perfect similarity
with the {\it Entwurf}, the attempt to rederive the field
equations from a variational principle and the investigation of
their compatibility.

The final fate of the approach is documented by a French version
of the {\it Annalen} article, three popular accounts of the
present state of field theory that mention distant parallelism as
a promising recent progress, as well as four further notes in the
{\it Sitzungsberichte}, two of them co-authored with Walther
Mayer.

\subsection{Improving the derivation of the field equations}
\label{sec:oldage1}

When the manuscripts for their {\it Annalen} papers were still
sitting with the publisher, Cartan and Einstein had occasion for a
personal encounter. In November 1929, Einstein travelled to Paris.
He was awarded an honorary doctorate and also gave two lectures at
the Institut Henri Poincar\'e.\footnote{The lectures were given on
8 and 12 November, and the awarding of the honorary doctorates
took place at the ceremony of the annual reopening of the Paris
university at the Sorbonne on 9 November, see
\cite[pp.~21f.]{Debever1979} for details.} Einstein's lectures at
the Institut Henri Poincar\'e were subsequently published in
French in the institute's {\it Annales} \cite{Einstein1930b}. This
French account of the theory closely parallels the version in the
{\it Mathematische Annalen}, being slightly more explicit in the
mathematical details.

The personal encounter between Einstein and Cartan also seemed to
have resulted in some further work of the latter on the theory.
This is witnessed by a few extensive and technical manuscripts
that have been published in the Einstein-Cartan correspondence
\cite{Debever1979}. One such manuscript
\cite[pp.~32--55]{Debever1979} by Cartan immediately led Einstein
to publish an improved version of the compatibility proof in his
{\it Annalen} paper, even before that paper was available in
print.\footnote{In a postscript to a letter to Cartan, dated 10
January, Einstein complains: ``It is remarkable that the {\it
Mathematische Annalen} has such terrible constipation that after,
so many months, it has not been able to excrete what it has
absorbed.'' \cite[p.121]{Debever1979}. The correspondence between
Einstein and Cartan at the end of 1929 was intense and it was
Cartan who took the lead by working on the mathematical side of
the problem. ``I am very fortunate that I have acquired you as a
coworker (Mit-Strebenden). For you have exactly that which I lack:
an enviable facility in mathematics.'' (18 December 1929). The
correspondence with Cartan on teleparallelism reminds of a similar
correspondence with Einstein and Levi-Civita on mathematical
details of the derivation of the {\it Entwurf} equations. Einstein
seems to have had comparable feelings of appreciation for
Levi-Civita to whom he wrote in 1917: ``It must be nice to ride
these fields on the cob of mathematics proper, while the likes of
us must trudge along on foot.''(``Ich bewundere die Eleganz Ihrer
Rechnungsweise. Es muss h\"ubsch sein, auf dem Gaul der
eigentlichen Mathematik durch diese Gefilde zu reiten, w\"ahrend
unsereiner sich zu Fuss durchhelfen muss.'') Einstein to
Levi-Civita, 2 August 1917, \cite[Doc.~368]{Einstein1998}.} On
December 12, 1929, Einstein submitted a communication to the
Prussian Academy on the ``Compatibility of the Field equations in
the Unified Field Theory'' \cite{Einstein1930f}. In this short
note, Einstein first gave a few critical remarks on his earlier
papers. These concerned the divergence operation introduced in
\cite{Einstein1929b} which Einstein now considered inappropriate
because it does not vanish when applied to the fundamental tensor.
Einstein also mentioned that the compatibility proof given in that
paper was untenable because it erroneously assumed the existence
of a set of identities for the field equations. Finally, Einstein
pointed out that his discussion of the magnetic field equation in
\cite{Einstein1929e} was based on an unjustified assumption.

The major part of the note, however, was devoted to a brief survey
of the mathematical apparatus of the theory (which Einstein
probably gave because the long review paper had not yet come out)
and a discussion of the compatibility issue. The main point was
that Einstein had learnt from Cartan that the compatibility proof
could be improved.\footnote{``Der Kompatibilit\"atsbeweis ist auf
Grund einer brieflichen Mitteilung, welche ich Hrn.~Cartan
verdanke [...], gegen\"uber der in den Mathematischen Annalen
gegebenen Darstellung etwas vereinfacht.''
\cite[p.~18]{Einstein1930f}.} The point was that the strange
identity (\ref{eq:identity7}) could, in fact, be substituted by
the simple identity
\begin{equation}
G^{\mu\alpha}_{;\mu} + \Lambda^{\alpha}_{\sigma\tau}G^{\sigma\tau}
\equiv 0. \label{eq:identity8}
\end{equation}
The compatibility proof was now given by Einstein for the field
equations (\ref{eq:fieldeq1}) and (\ref{eq:fieldeq2}) on the basis
of the identities (\ref{eq:Id2}), (\ref{eq:identity6}), and
(\ref{eq:identity8}).

The issue of proving compatibility was taken up again in a very
brief note from July 1930 \cite{Einstein1930g} where Einstein
again introduced a divergence operation $\dots_{/\alpha}$ and
showed that it may be used to prove the compatibility of certain
equations that are similar to his field equations. He did not,
however, discuss the consequences for his system of equations
(\ref{eq:fieldeq1}) and (\ref{eq:fieldeq2}) explicitly.

\subsection{Promoting and defending the theory}
\label{sec:oldage2}

In October 1929, Einstein was asked to substitute for the late
secretary of state Leipart to give a lecture to some 800 invited
members of the Kaiser-Wilhelm society and other representatives of
scientific and cultural institutions and ministries. Einstein
agreed and gave a talk on the {\it Problem of Space, Field, and
Ether in Physics} on December 11, 1929.\footnote{Harnack to
Einstein, 18 October 1929, EA~1-084.} Essentially the same talk
was delivered to a large audience on the opening day of the {\it
Second World Power Conference} which took place in Berlin from
16--25 June, 1930.\footnote{K\"ortgen to Einstein, 22 February
1930, EA~1-085.} The text of this lecture was then published in
the conference's {\it Transactions}
\cite{Einstein1930d}.\footnote{A similar popular account of {\it
Space, Ether and the Field in Physics} was published in {\it Forum
Philosophicum} \cite{Einstein1930c} together with an English
translation. Indeed, the text of the two penultimate paragraphs of
this version and \cite{Einstein1930d} that characterize the
distant parallelism are identical. A two-page abbreviated version
of \cite{Einstein1930c} also mentions the distant parallelism
approach \cite{Einstein1930e}.} Just as in the articles of the New
York and London Times, this lecture gave a historical account of
our concepts of space, starting with our prescientific notion,
discussing Euclidean geometry, Cartesian analytic geometry,
Newtonian absolute space, the ether concept of 19th-century
electrodynamics, special relativity, and Riemannian geometry of
general relativity. In the final paragraphs, Einstein hinted again
at the latest progress of a ``unitary field theory'' based on a
mathematical structure of space which is ``a natural
supplementation of the structure of space according to the
Riemannian metric.'' He explained again the meaning of distant
parallelism and wrote, a little less self-confident than in the
{\it Times}
\begin{quote}
For the mathematical expression of the field-laws we require the
simplest mathematical conditions to which such a structure of
space can conform. Such laws seem actually to have been discovered
and they agree with the empirically known laws of gravitation and
electricity in first approximation. Whether these field-laws will
also yield a usable theory of material particles and of motions
must be determined by deeper mathematical investigations.
\cite[p.~184]{Einstein1930c}
\end{quote}

Einstein also defended his new theory in private correspondence. A
succinct example is a rebuttal of a saucy criticism by Wolfgang
Pauli. With respect to the theory as presented in the {\it
Annalen}, Pauli wrote that he no longer believed that the quantum
theory might be an argument for the distant parallelism after Weyl
and Fock had shown that Dirac's electron theory could be
incorporated into a relativistic gravitation theory in a way that
is not {\it globally} but {\it locally} Lorentz covariant. Pauli
also wrote that he did not find the derivation of the field
equations convincing, complained that the Maxwell equations would
be obtained only in differentiated form, and expressed doubts
whether an energy-momentum tensor of the field could be found. He
finally missed the validity of the classical tests of general
relativity, perihelion motion and gravitational light bending.
Pauli concluded
\begin{quote}
I would take any bet with you that you will have given up the
whole distant parallelism at the latest within a year from now,
just as you had given up previously the affine theory. And I do
not want to rouse you to contradiction by continuing this letter,
so as not to delay the approach of the natural decease of the
distant parallelism theory.\footnote{``[...] ich w\"urde jede
Wette mit Ihnen eingehen, dass Sie sp\"atestens nach einem Jahr
den ganzen Fernparallelismus aufgegeben haben werden, so wie Sie
fr\"uher die Affintheorie aufgegeben haben. Und ich will Sie nicht
durch Fortsetzung dieses Briefes noch weiter zum Widerspruch
reizen, um das Herannahen dieses nat\"urlichen Endes der
Fernparallelismustheorie nicht zu verz\"ogern.'' Pauli to
Einstein, 19 December 1929, \cite[Doc.~239]{Pauli1979}.}
\end{quote}
Einstein found Pauli's critique ``amusing but a little
superficial.'' Without going into details, he argued that neither
Pauli was in a position to ``view the unity of the forces in
nature from the correct stand point'' and one may not discard his
theory before its mathematical consequences were thoroughly
thought through. He claimed
\begin{quote}
that with a deeper look at it you would certainly understand that
the system of equations advanced by me is forced by the underlying
structure of space, particularly since the compatibility proof of
the equations could be simplified in the meantime. Forget what you
have said and engross yourself in the problem with such an
attitude as though you had just come down from the moon and would
yet need to form a fresh opinion. And then don't utter an opinion
before at least a quarter of a year has passed.\footnote{``Dass
das von mir aufgestellte Gleichungssystem zu der zugrunde gelegten
Raumstruktur in einer zwangl\"aufigen Beziehung steht, w\"urden
Sie bei tieferem Studium bestimmt einsehen, zumal der
Kompatibilit\"atsbeweis der Gleichungen sich unterdessen noch hat
vereinfachen lassen. Vergessen Sie, was Sie gesagt haben und
vertiefen Sie sich einmal mit solcher Einstellung in das Problem,
wie wenn Sie soeben vom Mond heruntergekommen w\"aren und sich
erst frisch eine Meinung bilden m\"ussten. Und dann sagen Sie erst
etwas dar\"uber, wenn mindestens ein Vierteljahr vergangen ist.''
Einstein to Pauli, 24 December 1929, \cite[Doc.~240]{Pauli1979}.}
\end{quote}

\subsection{Elaboration of consequences}
\label{sec:oldage3}

Both the long review paper in the {\it Annalen} (as well as its
French counter part \cite{Einstein1930b}) and this short note end
with the expression of the next step along the teleparallel
approach:
\begin{quote}
The most important question that is now tied to the (rigorous)
field equations is the question of the existence of
singularity-free solutions which can represent electrons and
protons.\footnote{``Die wichtigste an die (strengen)
Feldgleichungen sich kn\"upfende Frage ist die nach der Existenz
singularit\"atsfreier L\"osungen, welche die Elektronen und
Protonen darstellen k\"onnen.'' \cite[p.~23]{Einstein1930f}.}
\end{quote}

This problem was indeed attacked by Einstein in his pursuit of the
teleparallel program. It was a problem where he found help by a
collaborator. With Grommer and M\"untz leaving for Minsk resp.\
Leningrad, Einstein may have found himself in need of new
collaborators. After contacting Richard von Mises about suitable
candidates, he was recommended Walther Mayer (1887--1948), then
{\it Privatdozent} for mathematics in Vienna.\footnote{Richard von
Mises to Einstein, 17 December 1929, EA~18-225. For biographical
information on Mayer, see \cite[pp.~492--494]{Pais1982}.} Mayer
was an expert in invariant theory and differential geometry.
Einstein was interested and technical arrangements were quickly
agreed upon. Mayer arrived in Berlin some time in January 1930 but
apparently began to work on problems associated with the
teleparallelism approach before his arrival.\footnote{See Einstein
to Mayer, 1 January 1930, EA~18-065.}

The collaboration with Mayer proved to be of immediate success.
Already on 20 February 1930, they presented a first joint paper
for publication in the Academy proceedings
\cite{EinsteinMayer1930}. In it they discussed two special
solutions for the teleparallel field equations, i.e.\ those
presented and derived in the {\it Annalen} paper
\cite{Einstein1930a}, the case of spatially spherical symmetry,
and the static case of an arbitrary number of non-moving,
non-charged mass points.

Assuming spatial rotation symmetry as well as reflection symmetry,
their solution explicitly read
\begin{alignat}{2}
h_s{}^{\alpha} &=
\frac{\delta_{s}{}^\alpha}{\sqrt[4]{1-\frac{e^2}{r^4}}}, \qquad
\alpha, s
= 1,2,3, & \qquad h_s{}^4 &= 0, \notag\\
h_4{}^{\alpha} &= \frac{e}{\sqrt[4]{1-\frac{e^2}{r^4}}}, \qquad
\alpha=1,2,3, & \qquad h_4{}^4 &=
1+m\int\sqrt[4]{1-\frac{e^2}{r^4}\frac{dr}{r^2}}
\end{alignat}
where $r^2 = \sum_{a=1}^3x^ax_a$ is the spatial distance from the
origin, and $e$ and $m$ two constants to be identified with the
charge and mass of the particle.

For vanishing charge $e$, the solution reduces to
\begin{equation}
h_s{}^{\alpha} = \delta_{s}{}^{\alpha} \qquad s = 1,2,3, \qquad
h_{4}{}^{\alpha} = \delta_4^{\alpha}\left(1 +
\sum_j\frac{m_j}{r_j}\right), \qquad m_j= \text{konst}.
\label{eq:solutionuncharged}
\end{equation}
Einstein and Mayer interpreted (\ref{eq:solutionuncharged}) to the
effect that two or more uncharged massive particles could stay at
rest with arbitrary distance from each other. They emphasized,
however, that the solution was singular and that the theory would
not allow to derive equations of motion for such singular
solutions. On the contrary, it must be demanded that only
non-singular solutions are interpreted as representing elementary
particles.

\subsection{The demise of the {\it Fernparallelismus} approach}
\label{sec:oldage4}

Roughly a decade later, Einstein summarized his reasons for
abandoning the distant parallelism approach
\begin{quote}
Today, I am firmly convinced that the distant parallelism does not
lead us to an acceptable representation of the physical field.
From the reasons for this I will only give two:

1) One cannot find a tensor-like representation of the
electromagnetic field.

2) The theory leaves too large a freedom for the choice of the
field equations.%
\footnote{``Ich bin heute fest davon \"uberzeugt, da{\ss} der
Fern-Parallelismus zu keiner brauchbaren Darstellung des
physikalischen Feldes f\"uhrt. Von den Gr\"unden will ich nur zwei
anf\"uhren.

1) Man gelangt nicht zu einer tensor-artigen Darstellung des
elektromagnetischen Feldes

2) Die Theorie l\"a{\ss}t eine zu gro{\ss}e Freiheit f\"ur die
Wahl der Feldgleichungen'' Einstein to Salzer, 29 August 1938
\cite[p.~90]{Salzer1974}.}
\end{quote}

I will not comment here on the first point mentioned by Einstein.
But the second point is, I believe, well illustrated by Einstein's
last paper on this approach. It is again a paper coauthored with
Mayer, and it is concerned with a ``systematic investigation of
compatible field equations that can be set in a Riemannian space
with distant parallelism'' \cite{EinsteinMayer1931}. The paper is
remarkable in two respects. For one, it was presented to the
Academy on 23 April, 1931, and hence appeared some nine months
later than the last two-page note from July 1930. All other papers
on the approach were published within at most six months in
between. Even in the pure chronology, the paper thus appears as a
belated and final word on the fate of the approach. Second, this
paper, as we will see, is a quite unusual paper for Einstein in
its technicality.

To discuss the admissible field equations, Einstein and Mayer
demand that these be linear in the second derivatives of the field
variables $h_{s\nu}$ and at most quadratic in the first
derivatives. They also argue that the identities which the left
sides of the field equations satisfy should contain these
variables only linearly and in first order, and they also should
contain the torsion tensor $\Lambda^{\alpha}_{\mu\nu}$ explicitly
only linearly. Using the notation of the previous papers, Einstein
and Mayer now make the following {\it ansatz} for the field
equations of the theory
\begin{equation}
0=G^{\mu\alpha} =
p\Lambda^{\alpha}_{\underline{\mu}\underline{\nu};\nu} +
q\Lambda^{\mu}_{\underline{\alpha}\underline{\nu};\nu} +
a_1\phi_{\underline{\mu};\alpha} +
a_2\phi_{\underline{\alpha};\mu} +
a_3g^{\mu\alpha}\phi_{\underline{\nu};\nu} +
R^{\mu\alpha} \label{eq:ansatz1}
\end{equation}
where $p$, $q$, $a_1, \dots a_3$ are arbitrary real coefficients,
and $R^{\mu\alpha}$ denotes an as yet unspecified term that is
quadratic in the $\Lambda$'s.

They also write the divergence identity that is to be satisfied in
the following general form
\begin{align}
0 \equiv G^{\mu\alpha}_{;\mu} + AG^{\mu\alpha}_{;\mu} +
G^{\sigma\tau}\left(
c_1\Lambda^{a}_{\sigma\tau} +
c_2\Lambda^{\underline{\tau}}_{\sigma\underline{a}} +
c_3\Lambda^{\underline{\sigma}}_{\tau\underline{a}}\right)
&+
c_4G^{\alpha\sigma}\phi_{\sigma} +
c_5G^{\sigma\alpha}\phi_{\sigma} \notag \\
&+ c_6G^{\underline{\sigma}\sigma}\phi_{\underline{a}} +
BG^{\underline{\sigma}\sigma}_{;\underline{a}},
\label{eq:ansatz2}
\end{align}
where again $A$, $c_1, \dots c_6$, and $B$ are unspecified
coefficients.

Einstein and Mayer explicitly admit the possibility of other terms
not contained in this ansatz, especially for $n=4$ dimensions.
Nevertheless they claim that the neglected terms would be rather
unnatural ones and that the general {\it ansatz} of
eqs.~(\ref{eq:ansatz1}), (\ref{eq:ansatz2}) is, in fact, the most
general one that is consistent with the restrictive conditions of
the problem.

Accepting the generality of the {\it ansatz}, the problem of
finding the manifold of admissible field equations then reduces to
the algebraic problem of determining the unspecified constants
$p$, $q$, $a_1, \dots a_3$, $A$, $c_1, \dots c_6$, and $B$, as
well as the constants implicitly contained in the generic term
$R^{\mu\alpha}$. This algebraic problem is straightforward but
formidable. One may well image that it took Einstein and Mayer a
while to find their way through the resulting explicit
equations.\footnote{A number of apparently related but otherwise
unidentified manuscript pages are extant in the Einstein Archives,
see, e.g., EA~62-003ff, EA~62-054ff, EA~62-132ff.} The result is
the subject of this final note on the {\it Fernparallelismus}
approach. Introducing a few simplifications, they nevertheless end
up with a system of 20 algebraic equations for 11 coefficients
which they list and discuss. Using a tree-like graphical
representation, they classify possible types of solutions which
they subsequently try to associate with known cases and solutions.

The final upshot of their investigation is summarized in the final
paragraph of their paper.
\begin{quote}
The result of the whole investigation is the following: In a space
with Riemann-metric and {\it Fernparallelismus} of the character
defined by (1), (2) [i.e.\ our eqs.~(\ref{eq:ansatz1}),
(\ref{eq:ansatz2})---TS] there are all in all four (nontrivial)
different types of (compatible) field equations. Two of these are
(non-trivial) generalizations of the original field equations of
gravitation, one of which is already known as resulting from a
Hamiltonian principle [cp.~(10) and (11)]. The remaining two types
are denoted in the paper by (13) and $\Pi_{221221}$.
\end{quote}
These are Einstein's final words in print on the {\it
Fernparallelismus} approach. The equations (10), (11), (13) of
their paper that they refer to and the expression $\Pi_{221221}$
indicate various field equations given in more or less explicit
form.

\section{Concluding remarks}
\label{sec:conclusion}

As indicated in the introduction and at various points along the
paper, the life cycle of the {\it Fernparallelismus} approach
shows a number of similarities with the life cycle of the {\it
Entwurf} theory of the years 1912--1915. For the sake of the
present account, I would like to recall the following features of
the fate of the {\it Entwurf} theory, the genesis, life, and
demise of which is well understood by recent historical
research.\footnote{For a historical account of the prehistory of
general relativity along the lines given here, see
\cite{RennSauer1999}. See also \cite{Norton1984},
\cite[ch.~V]{Stachel2002}, and \cite{Rennetalforthcoming} as well
as further references cited in these works.}

The theoretical framework of this theory crucially depended upon
the insight that the metric tensor field is the mathematical
ingredient needed to set up a generalized theory of relativity and
a theory of gravitation. This insight was made by Einstein some
time in the summer of 1912 but the mathematics associated with the
metric tensor field was not fully understood by Einstein in the
beginning. Somewhat fortuitously he was able to enter into an
intense collaboration with a befriended mathematician, Marcel
Grossmann, then his colleague at the Polytechnic in Zurich. The
subsequent development, as documented mainly by unpublished
manuscripts and correspondence, consisted in an intense search for
a gravitational field equation that would satisfy a number of
heuristic requirements. An analysis of Einstein's research notes
of that period showed that he pursued a dual strategy for finding
field equations. At one point, Einstein was content with a set of
field equations that was not generally covariant but seemed to
square best with most of his other heuristic requirements.
Einstein and Grossmann published their theory in their joint {\it
Entwurf}. The further development of this theory involved both the
elaboration of empirically relevant consequences, notably the
planetary perihelion anomaly, and the further mathematical
justification of its field equations, with particular emphasis on
the question of their uniqueness. By mid-1915 several difficulties
of the theory had become evident to Einstein, and it was then
abandoned in November 1915 and superseded by new, generally
covariant field equations, viz.\ the Einstein equations of today's
general relativity.

Reflecting on the ``biographical'' similarities between the {\it
Entwurf} theory and the {\it Fernparallelismus} theory, it seems
that there is a {\it systematic} reason for this
similarity.\footnote{I agree with the general thesis of
\cite{vanDongen2002} who identified methodological convictions for
Einstein's work on semi-vectors and on five-dimensional field
theory that had originated during the {\it Entwurf} period. In
contrast to van Dongen I would only emphasize the constraints and
inherent possibilities of the mathematical representation over the
role of explicit methodological reflections.} It resides in the
roles that the mathematical representation in terms of the metric
tensor field, resp.\ of the tetrad field and the search for field
equations for these quantities played in each theory.

In both cases, the mathematics associated with the new concept was
badly understood by Einstein in the beginning. In both cases, the
mathematics had been worked out before in purely mathematical
contexts. In both cases, it was through the mediation of more
mathematically trained colleagues that Einstein learnt about the
earlier relevant mathematical developments. More specifically, we
observe that after a relatively brief period where the
mathematical concepts of metric resp.\ tetrad were accepted as the
key elements, the further research program focussed on finding
field equations for these quantities. In the attempts to find,
derive, and justify those field equations, heuristic convictions
become visible that had been conceived in previous work.

In the case of the {\it Entwurf} theory, the relevant heuristic
assumptions could be identified as the equivalence hypothesis,
postulates of general covariance, energy-momentum conservation,
and of correspondence, i.e.\ the admissibility of the Newtonian
limit \cite{RennSauer1999}.

In the case of the {\it Fernparallelismus} approach, the
corresponding heuristic convictions still need to be identified
more precisely through the study of unpublished correspondence and
notes. It appears, however, that one may similarly identify a
number of postulates that play a similar role. Two such postulates
are the demands of distant parallelism and general covariance. We
also have a postulate that the known cases of the relativistic
gravitational field equation for vacuum and the Maxwell equations
shall be identifiable in some weak field limit. Third, we have
seen that Einstein postulated that non-singular, spatially
symmetric, stationary solutions can be found that can be
interpreted as elementary particles. Finally, he was postulating
that equations of motion should be derivable for those
particle-like solutions.

In the case of the {\it Entwurf} theory, the heuristic postulates
were mutually incompatible in Einstein's original understanding.
The incompatibility showed itself in Einstein's difficulty to find
field equations that would satisfy all four of his postulates at
the same time. As a consequence, Einstein developed a double
strategy of finding field equations that we have called the
mathematical resp.\ the physical strategy \cite{RennSauer1999}.
One strategy started from the postulates of general covariance and
tried to modify equations constructed on the basis of the Riemann
tensor in order to justify the more physically motivated
postulates of energy-momentum conservation and of obtaining the
Newtonian limit. The complementary strategy started from
expressions that guaranteed the Newtonian limit from the beginning
and tried to enlarge the covariance group so as to generalize the
relativity principle.

In the {\it Fernparallelimus} approach something similar seems to
be observable. Here again, we may distinguish two distinct
approaches to the problem of finding field equations. A
mathematical, variational approach started from a mathematically
well-defined {\it ansatz} but the problem was to obtain the
gravitational and electromagnetic field equations in first
approximation. The complementary physical strategy, the
overdetermination approach, on the other hand, started from
identities that guaranteed the validity of the gravitational and
electromagnetic equations from the outset. The drawback here was
the mathematical problem of proving the compatibility of the field
equations. In both cases, at the mature stage, Einstein settled
for the more physical approach.

Can we also compare the demise of the two theories? From the more
global perspective of Einstein's heuristics, the result of the
final paper \cite{EinsteinMayer1931} may be phrased as follows.
The overdetermination approach to finding field equations within
the distant parallelism framework had provided a manifold of
different admissible equations. These were not only difficult to
find and handle in their algebraic complexity. The approach also
seemed to encompass the equations produced by the alternative
variational approach and to produce even more admissible field
equations than that method.

In the case of the {\it Entwurf} theory of gravitation, several
difficulties accumulated before its demise. But what sealed the
fate of the {\it Entwurf} in the end was the success of its
alternative, the generally covariant Einstein equations
\cite[pp.~115ff]{RennSauer1999}. These equations gave the correct
value for the anomaly of the perihelion motion for Mercury and
they solved the energy-momentum problem by virtue of the
contracted Bianchi identities.

More than one reason was presumably responsible for Einstein's
loss of faith in the distant parallelism approach. The mere
algebraic complexity can hardly have been the decisive reason for
giving it up, certainly not from a logical point of view. But it
may have motivated Einstein to explore alternatives. More
problematic must have been the apparent impossibility to justify a
set of field equations uniquely. But here again it is hard to see
how this difficulty could be turned into a logically compelling
argument for giving up the approach. After all one could always
add new heuristic requirements, or justify particular equations
{\it post hoc} as it were by their subsequent success. But just as
in the case of the {\it Entwurf}, the final demise may have been
effected by the success of a different theory.

Indeed, only a few months later Einstein and Mayer presented a new
approach towards a unified theory \cite{EinsteinMayer1931a} that
may have seemed more promising to them at the time. In this
approach, the introduction of an independent orthonormal basis
field in some vector spaces associated with each point of the
manifold is again the crucial mathematical ingredient. But now the
frame fields and hence the vector spaces were no longer assumed to
be of the same dimension as the underlying manifold and hence they
were no longer to be identified with the tangent bundle. They were
now taken to be five-dimensional. The introduction of a
five-dimensional frame bundle pointed to a reconsideration of the
Kaluza-Klein approach. Since the underlying space-time manifold
was still assumed to be four-dimensional, the new approach was
also sufficiently different from earlier consideration of the
five-dimensional field theory that earlier arguments against the
Kaluza-Klein approach were no longer valid. Indeed, the
five-dimensional vector spaces may have seemed promising enough to
justify the abandoning of the {\it Fernparallelismus} approach for
the time being. In contrast to other approaches in his quest for a
unified theory, it seemed to have been a final demise, too.
Einstein apparently did not return to an exploration of the
conceptual framework of distant parallelism in his subsequent
quest for a unified field theory of gravitation and
electromagnetism.

\section*{Acknowledgments}

A preliminary version of this paper was presented at a conference
on the history of modern mathematics, held at the Open University,
Milton Keynes, in September 2002. I wish to thank Jeremy Gray for
the invitation to this meeting. I am also grateful to Walter
Hunziker and the Institute for Theoretical Physics at the ETH
Zurich for its hospitality during the summer 2002. I wish to thank
Jeroen van Dongen, Friedrich-Wilhelm Hehl, and Erhard Scholz for
some helpful comments on an earlier draft of this paper.
Unpublished correspondence by Einstein is quoted by kind
permission of the Albert Einstein Archives, The Hebrew University
of Jerusalem.

\section*{Appendix: A note on notation}

During the elaboration of the teleparallelism approach Einstein
introduced---and dropped---a few notational idiosyncrasies. For a
systematic reconstruction of the theory, these notational changes
are awkward to deal with. However, for a historical reconstruction
they provide very useful information. They help to identify and
date calculational manuscripts and they may provide clues as to
Einstein's reception of literature as well as to his heuristics.

I will summarize here the three notational pecularities associated
with the {\it Fernparallelismus} approach. They concern a) the
notation of the anholonomic indices of the tetrads, b) a ``new''
divergence operation, and c) a peculiar way of indicating raising
and lowering of indices.

Einstein rather consistently denotes the anholonomic indices ({\it
Bein-Indizes}) of the tetrads by latin indices and the holonomic
indices ({\it Koordinaten-Indizes}) by greek indices. As discussed
above in sec.~\ref{sec:childhood2}, Weitzenb\"ock had written to
Einstein shortly after the publication of Einstein's first two
notes on teleparallelism pointing out his priority with respect to
the Weitzenb\"ock connection. In Einstein's next publications, in
the Stodola-Festschrift \cite{Einstein1929a} and in
\cite{Einstein1929b} he already used Weitzenboeck's notation of
putting the anholonomic index to the left of the tetrad symbol:
$^sh_{\mu}$ with explicit reference to Weitzenb\"ock's paper. The
notation is used again, but for the last time in March 1929 in
\cite{Einstein1929e}. The {\it Annalen} paper of summer 1929
reverts to the previous right hand side notation. The left hand
side notation therefore should give a fairly accurate hint to
material dating between summer 1928 and summer 1929.

In his note from January 1929 \cite{Einstein1929b}, Einstein
introduced what he called a ``divergence'' of a tensor density
$\mathfrak{A} \equiv hA$, $h\equiv\operatorname{det}(h_{s\mu})$ by
the following definition:
\begin{equation}
\mathfrak{A}^{\sigma\cdot\cdot i}_{\tau\cdot\cdot/i} =
\mathfrak{A}^{\sigma\cdot\cdot i}_{\tau\cdot\cdot,i} +
\mathfrak{A}^{\alpha\cdot\cdot i}_{\tau\cdot\cdot}
    \Delta^{\sigma}_{\alpha i} + \cdots -
\mathfrak{A}^{\sigma\cdot\cdot i}_{\alpha\cdot\cdot}
    \Delta^{\alpha}_{\tau i} - \cdots .
\end{equation}
Here a subscript comma denotes ordinary coordinate differentiation
and the dots indicate further contravariant and covariant indices.
The new ``divergence'' coincides with the usual covariant
divergence $\mathfrak{A}^{\cdot\cdot\sigma}_{\cdot\cdot;\sigma}$
formed using the covariant derivative associated with the
Weitzenb\"ock connection $\Delta$, see (\ref{eq:Weitzconn}), for
the case of vanishing torsion:
\begin{equation}
\mathfrak{A}^{\cdot\cdot\sigma}_{\cdot\cdot;\sigma} \equiv
\mathfrak{A}^{\cdot\cdot\sigma}_{\cdot\cdot/\sigma} +
\mathfrak{A}^{\cdot\cdot\sigma}_{\cdot\cdot}\Lambda^{\sigma}_{\alpha\sigma}.
\end{equation}
Heuristically, it was introduced in the context of introducing the
overdetermination approach because the relevant identities take on
a compact form using this notation. Einstein used this notation
again in his note from March 1929 in which he goes back to the
Hamilton approach. However, in the {\it Annalen} paper, he
explicitly wrote that he no longer recognized a specific physical
meaning of that divergence operation.\footnote{``In fr\"uheren
Arbeiten habe ich noch andere Divergenzoperatoren eingef\"uhrt,
bin aber davon abgekommen, jenen Operatoren eine besondere
Bedeutung zuzuschreiben.'' \cite[p.~689]{Einstein1930a}.}

Strangely enough, Einstein did revert to this non-standard
divergence another time. In his short, two-page note
\cite{Einstein1930g} he reintroduced the divergence symbol for an
arbitrary tensor $A^{\nu}$
\begin{equation}
A^{\nu}_{/\nu} = A^{\nu}_{;\nu}- A^{\nu}\varphi_{\nu},
\end{equation}
where $\varphi_{\nu}\equiv\Lambda^{\alpha}_{\sigma\alpha}$. It is
also used, albeit rather inconspicuously, in two equations in
\cite{EinsteinMayer1931}.

The third notational idiosyncrasy was also introduced in the
January 1929 note and was used in all subsequent papers on {\it
Fernparallelismus}.
\begin{quote}
Sometimes I will indicate the raising resp.\ lowering of an index
by underlining the corresponding index.\footnote{``Ich will
manchmal das Heraufziehen bzw.\ Hinunterziehen eines Index dadurch
andeuten, da{\ss} ich den betreffenden Index unterstreiche.''
\cite[p.~3]{Einstein1929b}.}
\end{quote}
An explicit example is (cp.~\cite[p.~693]{Einstein1930a})
\begin{align}
\Lambda^{\;\alpha}_{\underline{\mu}\;\underline{\nu}} &\equiv
\Lambda{\;\alpha}_{\beta\;\gamma}g^{\mu\beta}g^{\nu\gamma},\notag\\
\Lambda^{\;\underline{\alpha}}_{\mu\;\nu} &\equiv
\Lambda^{\;\beta}_{\mu\;\nu}g_{\alpha\beta}. \label{eq:raiseindex}
\end{align}


\end{document}